\documentclass[9pt,twocolumn,twoside]{pnas-new}
\templatetype{pnasresearcharticle} 

\newcommand{\red}[1]{\textcolor{black}{#1}}

\usepackage{pdfpages}

\begin{document}

\title{Tidally Induced Turbulence in the Abyssal Ocean}

\author[a,1]{Yidongfang Si}
\author[a]{Raffaele Ferrari}
\author[b]{Gunnar Voet}

\affil[a]{Department of Earth, Atmospheric, and Planetary Sciences, Massachusetts Institute of Technology, Cambridge, MA, USA}
\affil[b]{Scripps Institution of Oceanography, University of California San Diego, La Jolla, CA, USA.}

\leadauthor{Si, Ferrari and Voet}

\significancestatement{
Mixing along the ocean's sloping topography is a key driver of the abyssal upwelling limb of the global Meridional Overturning Circulation and its associated heat and carbon transport. Recent field campaigns in a North Atlantic submarine canyon have provided direct evidence of vigorous mixing producing upslope flows at rates of $\mathcal{O}(100)$\,m/day--the largest upslope flows ever recorded in the abyssal ocean. Integrating observations and numerical simulations, this study uncovers the physics driving the strong mixing. The mixing results from an instability of the oscillatory tidal flow, sloshing waters back and forth along the canyon. This physics may be relevant for many other canyons in the global ocean and, therefore, for our understanding of the abyssal ocean circulation.
}

\authorcontributions{Conceptualization: R.F. and Y.S. Methodology: R.F. and Y.S. Investigation: Y.S. Visualization: Y.S. Supervision: R.F. Writing: Y.S. and R.F. Observational data: G.V.}
\authordeclaration{The authors declare no competing interests.}
\correspondingauthor{\textsuperscript{1}To whom correspondence should be addressed. E-mail: y\_si@mit.edu}


\begin{abstract}
It has been suggested that the upwelling branch of the abyssal overturning circulation is characterized by strong flows driven by turbulence along sloping topography. The Boundary Layer Turbulence field campaign has provided direct evidence for strong upslope flows along a deep submarine canyon of the Rockall Trough. Turbulent overturning events spanning 200~m in the vertical were observed every tidal cycle, suggesting that the strong tidal flows in the canyon periodically undergo some form of instability. However, it is shown that the flow never satisfied the classical instability condition for time-independent sheared flows in a stratified fluid commonly used in oceanographic studies to determine transition to turbulence. This study illustrates that the time dependence of the tidal flow changes the stability properties and explains the observed transition to a turbulent state. The findings suggest that turbulent mixing induced by oscillatory shear flow may be ubiquitous over sloping topography and play an important role in deep ocean mixing that supports the global overturning circulation.
\end{abstract}

\dates{This manuscript was compiled on \today}
\doi{\url{www.pnas.org/cgi/doi/10.1073/pnas.XXXXXXXXXX}}

\maketitle
\thispagestyle{firststyle}
\ifthenelse{\boolean{shortarticle}}{\ifthenelse{\boolean{singlecolumn}}{\abscontentformatted}{\abscontent}}{}

\firstpage[15]{3}




Abyssal mixing plays a crucial role in the deep branch of the Meridional Overturning Circulation (MOC). This circulation strongly influences Earth's climate by redistributing heat, salt, nutrients, and carbon throughout the globe \citep{marshall2012closure,talley2013closure}.
The textbook description of the deep branch of the MOC is that dense water sinks to the ocean bottom at high latitudes and gradually upwells toward the sea surface in the ocean interior~\citep{munk1966abyssal}. 
However, in recent decades, observations and theoretical studies have suggested that the upwelling is confined to thin boundary layers along the sloping seafloor, specifically along ridges, seamounts, and continental slopes where turbulent mixing is particularly intense~\citep{polzin1997spatial,ferrari2016turning,mcdougall2017abyssal}. There has been much literature on abyssal mixing along ridges and seamounts with mild slopes~\citep{toole1997near,ledwell2000evidence,laurent2001buoyancy,laurent2012turbulence,polzin2009abyssal,ruan2020mixing,drake2022dynamics}, while less is known about the fluid dynamics that support mixing along steep continental slopes. Given that steep continental shelves surround all oceans, they may be a major conduit of water upwelling toward the surface if they support vigorous mixing.

Turbulent mixing above mildly sloping topography is mainly supported by internal wave breaking~\citep{waterhouse2014global}, which is largely generated by barotropic tides interacting with underwater topography~\citep{garrett2007internal,lamb2014internal,zhang2008resonant,gayen2011direct}. The main exception is the Southern Ocean, where the interaction between the strong geostrophic eddy field and the abyssal topography is an additional important source of internal waves~\citep{nikurashin2014impact,scott2011global}. 
Observational campaigns and theoretical studies have confirmed that tidal flows over seamounts and ridges with mild slopes result in the radiation of long internal waves. The long waves propagate a long distance before they scatter all their energy into shorter waves through nonlinear wave-wave interaction~\citep{staquet2002internal}. Ultimately, the shorter waves become nonlinear enough to break several hundred meters above the topography~\citep{polzin1997spatial,st2002estimating}.

\begin{figure*}[!ht]
\centering
\includegraphics[width=0.88\linewidth]{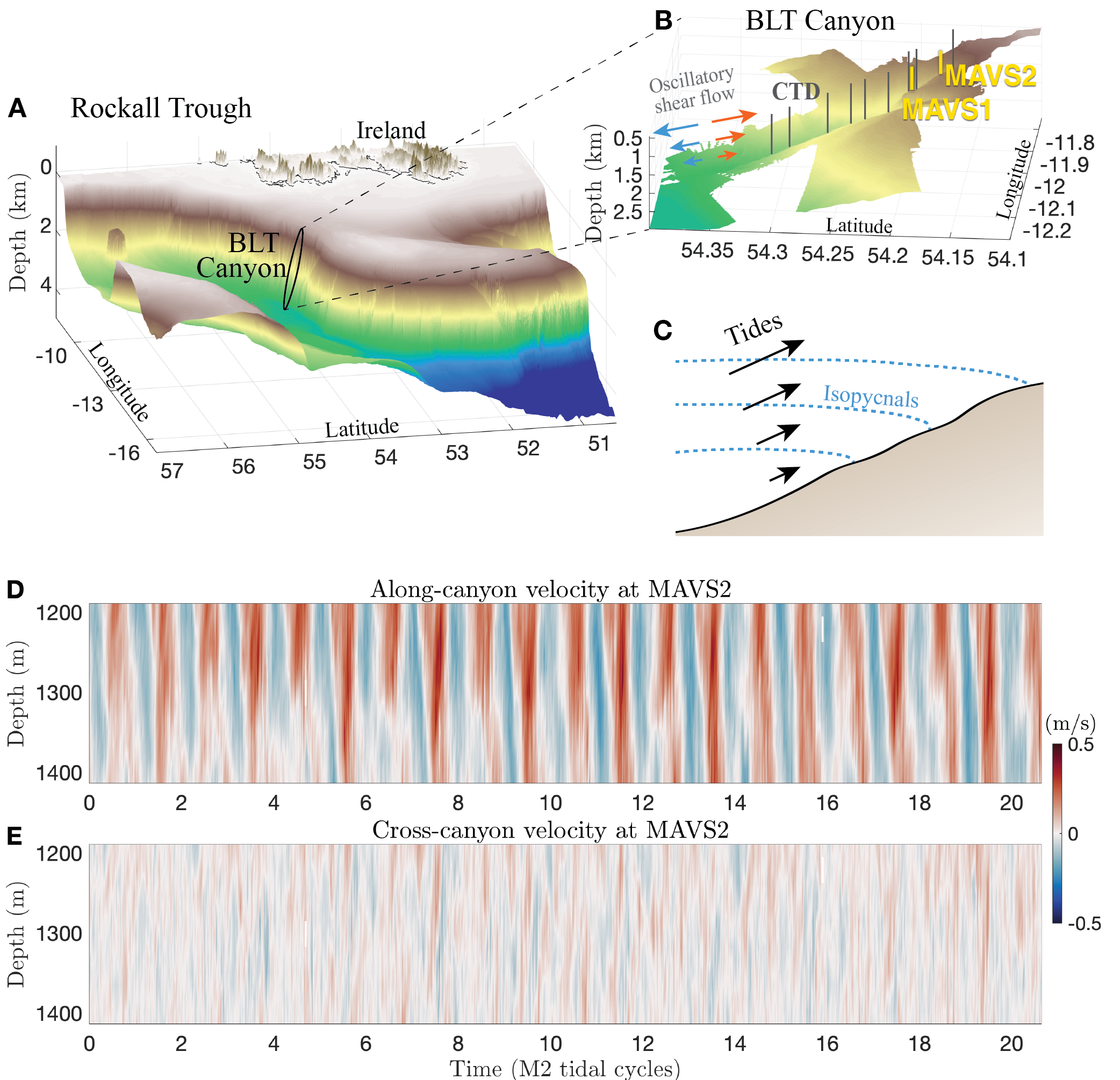}
\caption{
Observations from the Bottom Boundary Layer Turbulence (BLT) field campaign.
(\textit{A}) The bathymetry of the Rockall Trough from the General Bathymetric Chart of the Oceans (GEBCO) \citep{GEBCO}. 
(\textit{B}) The canyon bathymetry measured by a multibeam echo sounder during the BLT campaign \citep{voet24bathymetry}. 
The ellipse in panel (\textit{A}) marks the canyon's location where the measurements were made.
The gray and yellow bars denote the Conductivity--Temperature--Depth (CTD) stations and two moorings equipped with Modular Acoustic Velocity Sensors (MAVS), respectively.
The red and blue arrows represent the positive and negative shear phases.
(\textit{C}) Schematic illustrating the impact of shear on water density. 
During the positive shear phase, the tidal velocity shear advects denser fluid over less dense fluid, reducing the density stratification.
(\textit{D}-\textit{E}) Along-canyon and cross-canyon velocities observed by the MAVS2 mooring \cite{voet24moored} as a function of depth and time. The horizontal axes indicate M2 tidal cycles (since 2\,p.m. on 2021-07-07).}
\label{pnas_fig1}
\end{figure*}

Recently, the Bottom Boundary Layer Turbulence and Abyssal Recipes (BLT) field campaign has provided compelling evidence of significant turbulence in a steep canyon experiencing strong tidal flows along the eastern margin of the Rockall Trough in the Northeast Atlantic (Fig.~\ref{pnas_fig1}). The BLT canyon was chosen as a natural laboratory to investigate turbulent mixing and diapycnal (\textit{i.e.,} across density surfaces) upwelling along steep topography~\citep{garabato21cruise,voet24moored,van2024near}. Microstructure probes documented bursts of energetic turbulence and mixing spanning 200\,m above the seafloor and lasting a few hours every tidal cycle (Fig.~\ref{pnas_fig2})~\cite{garabato2024convective,alford2025buoyancy}.
A dye release experiment provided direct evidence of a rapid diapyncal flow up to 100\,m/day~\citep{wynne2024observations}, likely associated with the turbulence bursts. These observations suggest that turbulence and mixing are more confined to the seafloor when occurring along steep rather than mild topographic slopes. This study investigates what type of instability triggers turbulence in a steep canyon experiencing a tidal flow, such as the canyon sampled in the BLT campaign and others commonly found along continental shelves.

The BLT observations indicated that, averaged over a complete tidal period, the waters were stably stratified, \textit{i.e.}, density decreased with height above the bottom. The tidal flow was aligned along the canyon axis, and it had a vertical shear, \textit{i.e.,} it varied in the vertical (Fig.~\ref{pnas_fig1}\textit{D}-\textit{E}), as revealed by Acoustic Doppler Profilers~\citep{voet24moored}. Differential advection by this tidal flow reduced ocean stratification when the shear was positive (\textit{i.e.}, when the along-canyon velocity increased upward, Fig.~\ref{pnas_fig1}\textit{C}) and increased ocean stratification when the shear was negative. 
The turbulence bursts occurred when the shear was positive. Two primary instabilities are often associated with such a flow configuration. (i) The positive shear can be so large as to advect denser water from deeper in the canyon over lighter waters along the seafloor as illustrated in Fig.~\ref{pnas_fig1}\textit{C}. This scenario leads to a so-called convective instability where the denser fluid collapses toward the sea floor under gravity~\citep{gayen2011boundary}.
(ii) Even if the shear is not sufficiently strong to overturn the stratification, a second form of instability, known as the Kelvin--Helmholtz~(KH) instability~\cite{vallis2017atmospheric}, can develop. KH instability develops if the kinetic energy associated with the sheared velocity is large enough to \red{overcome the stabilizing effect of} the density stratification. For a time-independent shear, away from solid boundaries, this occurs when the shear squared, $|\partial_{\tilde z} {\bf u}|^2$, where $\bf u$ is the horizontal velocity and $\partial_{\tilde z}$ is the vertical derivative, is four times larger than the vertical stratification, defined as $N^2 =\partial_{\tilde z}\mathcal{B}$, where $\mathcal{B}=-g\rho/\rho_0$ is buoyancy, $g$ is gravitational acceleration, $\rho$ is density, and $\rho_0$ is a reference density. In fluid dynamics, this is quantified in terms of the Richardson number, $Ri=N^2 / |\partial_{\tilde z} {\bf u}|^2$, being smaller than 1/4. As shown below, in the BLT canyon, the shear associated with the large-scale background tide is never large enough to trigger turbulent overturns through either KH or convective instability.

Another form of instability, {\it parametric instability}, can develop when the sheared flow varies in time, like for the BLT tidal flow. Many of us have experienced parametric instability during our childhood, when we pumped our legs to make the swing go as high as possible. By lifting and lowering our legs, we varied the distance between the center of gravity and the suspension point, thereby changing the effective length of the swing. This modulation affected the swing's natural frequency. By adjusting our movements to match the swing's frequency, we induced a gradual increase in amplitude through parametric resonance. This parametric instability allowed our oscillations to grow without any external push, solely by synchronizing our body's motion with the swing's natural dynamics. A similar instability occurs when the tidal shear frequency (\red{or its rational multiples}) matches the natural frequency of oscillations of a water parcel~\citep{kelly1965stability,flierl2007nonlinear}. Demonstrating how this instability arises in a deep ocean canyon and leads to turbulence is the focus of the rest of this study.

In Sec.~\ref{sec:obs}, we demonstrate that the observed large-scale shear does not satisfy the necessary criteria to generate turbulent motions by either convective or KH instability. However in Sec.~\ref{sec:gcm}, we show that the tidal shear does go unstable using a high-resolution, non-hydrostatic, tidal-resolving model. We then use theory to demonstrate that the growth rate of the perturbations is consistent with a parametric instability.
In Sec.~\ref{sec:conclusion}, we summarize the results and discuss implications for the deep ocean circulation.

\section{\red{Analysis of Observations}}
\label{sec:obs}

\begin{figure*}[t!]
\centering
\includegraphics[width=0.9\linewidth]{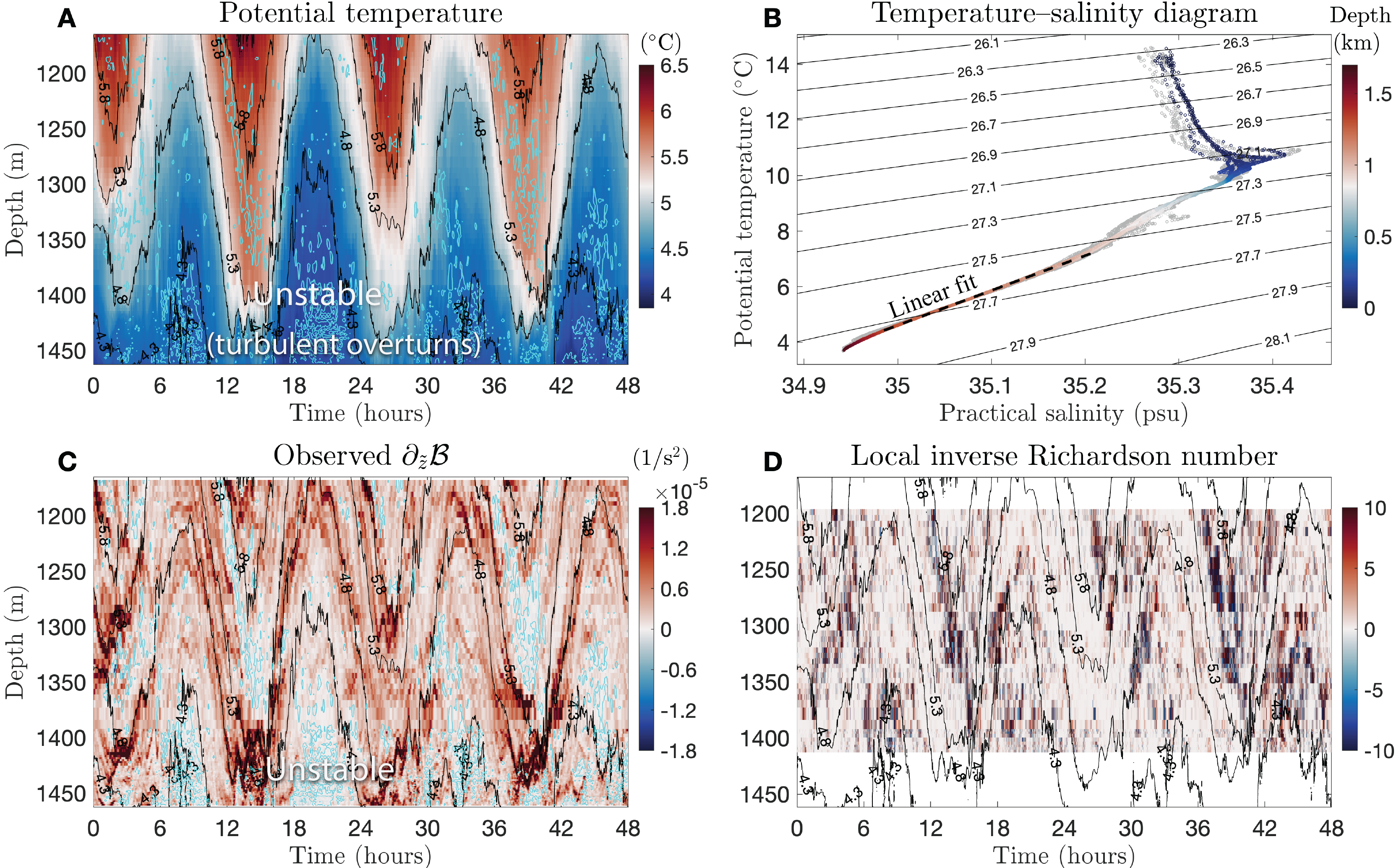}
\caption{
Observations from CTD stations and MAVS moorings.
(\textit{A}) Potential temperature at MAVS2 (see Fig.~\ref{pnas_fig1}\textit{B}). 
The black contours indicate temperature from 4.3$^\circ$C to 6.3$^\circ$C with an interval of 0.5$^\circ$C.
The cyan contours indicate locations with unstable stratification (vertical buoyancy gradient $\partial_{\tilde z} \mathcal B<0$). A Gaussian filter with a 15-min window has been applied to the temperature data to filter out high-frequency noise in the vertical buoyancy gradient $\partial_{\tilde z} \mathcal B$.
(\textit{B}) Temperature--salinity diagram from CTD profiles. The colored dots denote measurements from the CTD stations close to the MAVS2 mooring, with color indicating depth. The gray dots indicate measurements from the other 8 CTD stations throughout the canyon (Fig.~\ref{pnas_fig1}\textit{B}). The thin black contours show potential density minus 1000 in kg/m$^3$. (Potential density is the density of seawater after subtracting dynamically irrelevant compressive effects.) The black dashed line is a linear fit to the T--S diagram from 1050 to 1550\,m, covering the depth range of the MAVS moorings: $S=0.079 \,\,(\mathrm{psu}/^\circ\mathrm{C})\times T\,+34.64\,(\mathrm{psu})$. We use this linear fit to estimate salinity from temperature measured by the MAVS moorings.
(\textit{C}) Vertical buoyancy gradient ($\partial_{\tilde z} \mathcal B$) at MAVS2.
The cyan contours indicate locations with unstable stratification ($\partial_{\tilde z} \mathcal B<0$).
(\textit{D}) The local inverse Richardson number, $Ri^{-1}$, estimated from velocity data (Fig.~\ref{pnas_fig3}\textit{A}) linearly interpolated onto the temperature data grid.
The horizontal axes of panels (\textit{A},\,\textit{C},\,\textit{D}) indicate hours since 12\,a.m. on 2021-07-18.
}
\label{pnas_fig2}
\end{figure*}

\begin{figure*}[t!]
\centering
\includegraphics[width=0.88\linewidth]{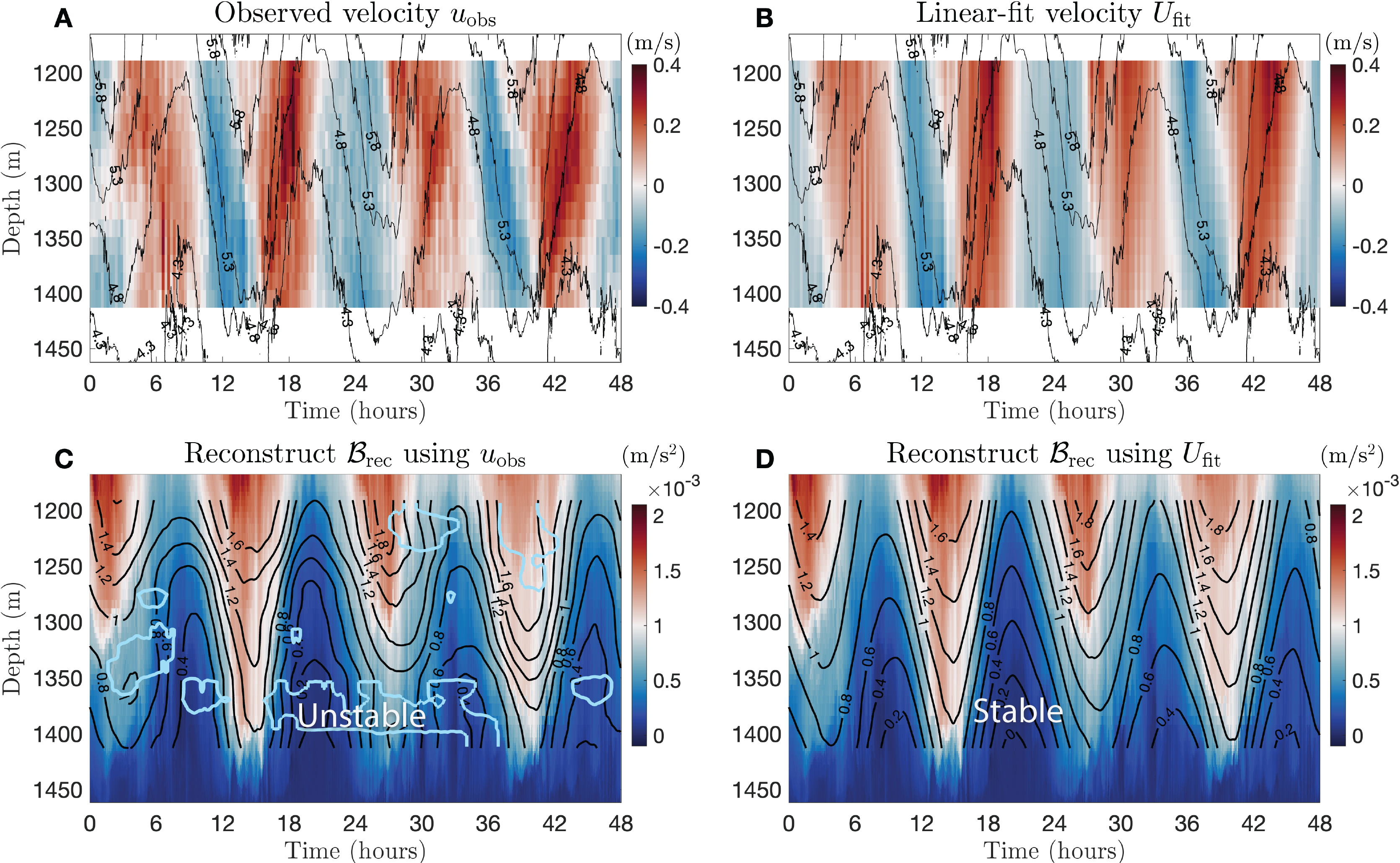}
\caption{
(\textit{A}--\textit{B}) Observed velocity ($u_\mathrm{obs}$) and linear-fit velocity  ($U_\mathrm{fit}$) at MAVS2.
The black contours indicate potential temperature from 4.3$^\circ$C to 6.3$^\circ$C with an interval of 0.5$^\circ$C.
(\textit{C}) \red{The color shading represents the observed buoyancy $\mathcal{B}$. The black contours represent the reconstructed buoyancy $\mathcal B_\mathrm{rec}\times1000$} using the kinematic model (Eq.~\ref{eq:reconstruct B}) with observed temperature and along-canyon-slope velocity ($u_\mathrm{obs}$), and an initial condition of $\partial_{\tilde z}\mathcal{B}_\mathrm{rec}\big\vert_{t=0}=N^2$, where $N $ is the vertical- and time-averaged buoyancy frequency. The cyan contours indicate locations with $\partial_{\tilde z} \mathcal B_\mathrm{rec} <0$ (unstable stratification).
(\textit{D}) Same as panel \textit{C}, but using the \textit{linear-fit} velocity ($U_\mathrm{fit}$) in the kinematic model.
The horizontal axes of all panels indicate hours since 12 a.m. on 2021-07-18.}
\label{pnas_fig3}
\end{figure*}

The BLT campaign surveyed the canyon's background hydrography with 9 profiles collected with a Conductivity--Temperature--Depth (CTD) Sensor along the canyon thalweg (Fig.~\ref{pnas_fig1}\textit{B}).
High-frequency measurements of temperature, velocity, dissipation rate of temperature variance (to quantify mixing rates), and dye concentration (used in a dye release experiment) were collected via moorings and a fastCTD~\cite{wynne2024observations,voet24moored}. (A fastCTD is a vertical profiler that measures conductivity, temperature, and pressure from which one can infer salinity, temperature, and depth.)
In this study, we use data from the CTDs and two moorings (MAVS1 and MAVS2).
Temperature was measured by sensors along the moorings every second, with a vertical resolution ranging from $\sim$1\,m near the bottom to $\sim$5\,m at 300\,m above the bottom (Fig.~\ref{pnas_fig2}\textit{A}). 
While salinity was not measured along the moorings, CTD data reveal a linear relationship between potential temperature and salinity within 1000$\sim$1500 m depth range (Fig.~\ref{pnas_fig2}\textit{B}), which can be used to reconstruct salinity profiles from temperature profiles. This allows us to estimate density and buoyancy frequency from the high-frequency temperature data recorded by the moorings, using the seawater equation of state which depends on temperature, salinity, and pressure/depth (Fig.~\ref{pnas_fig2}\textit{C}) \citep{vallis2017atmospheric}. 
Additionally, tidal velocities were recorded every 15 minutes with an Acoustic Doppler Current Profiler, with a vertical grid resolution 16\,m
(Figs.~\ref{pnas_fig1}\textit{D}-\textit{E}, \ref{pnas_fig3}\textit{A}).

The cyan contours in Fig.~\ref{pnas_fig2}(\textit{A},\textit{C}) highlight areas with unstable stratification (\textit{i.e.}, negative vertical buoyancy gradient), revealing deep overturning events every tidal cycle, spanning the full $\sim$200\,m in the vertical.
Next, we examine whether the density overturns can be attributed to either convective instability or KH instability.

\vspace{10pt}
\noindent (1) \textbf{Test for the Convective Instability Hypothesis}
\vspace{5pt}

We rotate the observed velocities to align with the canyon slope as the flow closely followed topography (Fig.~\ref{pnas_fig1}):
\begin{equation}
    u_\mathrm{obs}\!=\!\tilde u\cos\theta\! +\! \tilde w\sin\theta,
\end{equation}
where $\tilde u$ is the horizontal velocity along the canyon thalweg, $\tilde w$ is the vertical velocity, and $\theta$ is the topographic slope.
The tilde symbol $\tilde \cdot$ denotes quantities in the normal horizontal-vertical coordinate $(\tilde x, \tilde z)$, differentiating them from those in the slope-aligned coordinate $(x,z)$. 
Only the along-canyon velocity is considered, as the canyon walls strongly suppressed the cross-canyon flow (Fig.~\ref{pnas_fig1}\textit{D}-\textit{E}).

The observed shear is characterized by a smooth tidal flow together with small-scale fluctuations associated with turbulence. Ma et al.~\cite{ma2025tide} show that the tidal flow in the canyon is associated with a standing internal wave whose vertical scale is set by the canyon's geometry, including depth, length, and topographic slope. At the location of both MAVS1 and MAVS2, the strongest shear is confined to the bottom half of the canyon or about 200-300 meters. To extract the background shear associated with the tide, we perform a vertical linear fit to the observed along-canyon velocity over the full depth of MAVS2 measurements ($\sim$224\,m), $u_\mathrm{obs}(t,z)$, every 15 minutes:
\begin{equation}
    U_\mathrm{fit}(t,z) = u_\mathrm{bot}(t)+\Lambda_\mathrm{obs}(t)z,
    \label{eq:Ufit}
\end{equation}
where $u_\mathrm{bot}(t)=u_\mathrm{obs}(t,z=0)$ is the bottom velocity and $\Lambda_\mathrm{obs}(t)$ is the observed background velocity shear. 
Fig.~\ref{pnas_fig3}\textit{A}-\textit{B} demonstrates that the linear-fit method captures the pattern of the background velocity quite well.

It is useful to decompose the total buoyancy $\mathcal B (x,z,t)$ into a background component and a time-varying component arising from advection by the background tidal shear. The background buoyancy is defined as the time-mean field varying only in the vertical,
\begin{equation}
    B_0 = N^2\tilde z = N^2 (z\cos\theta+x\sin\theta),
\end{equation}
where $N $ is the vertical- and time-averaged buoyancy frequency. The time-dependent component, $b(z,t)$, arises because the tidal flow lifts the otherwise flat background density surfaces up and down along the sloping canyon. It depends only on $z$ and $t$, because we assume that the tidal velocity $u$ is parallel to the slope and depends on $z$ and $t$, but not on $x$. Thus,  
\begin{equation}
    \partial_t b \approx  - u\partial_x B_0 = - uN^2\sin\theta.
    \label{eq:dbdt}
\end{equation}
The along-canyon tidal flow $u$ can be either the full velocity, $u_\mathrm{obs}$, or the background one, $U_\mathrm{fit}$. The along-canyon buoyancy gradient is given by the projection of the vertical buoyancy gradient along the topographic slope, $\partial_x \mathcal B \approx \partial_x B_0 = N^2\sin\theta$, a relationship supported by the BLT observations (Fig.~3.16 in \citealp{Bethan_thesis}).

To determine whether the background tidal shear is sufficient to advect dense water over light water and trigger convective instability, we reconstruct the buoyancy field ($\mathcal{B}_\mathrm{rec}$) by integrating Eq.~\ref{eq:dbdt} over several tidal cycles, 
\begin{equation}
   \mathcal B_\mathrm{rec}(x,z,t)\approx  \mathcal B_\mathrm{rec}(x,z,t=0) -N^2 \sin\theta\int_{t=0}^t  u(z,t')\,\, {\rm d}t',
    \label{eq:reconstruct B}
\end{equation}
using either the observed along-canyon velocity ($u=u_\mathrm{obs}$) or its corresponding linear fit ($u=U_\mathrm{fit}$).
Here, $\mathcal B_\mathrm{rec}(x,z,t=0)=B_0$ is the initial condition.
To determine if the tidal shear is sufficient to drive overturns, we compute the vertical gradient of the reconstructed buoyancy ($\partial_{\tilde z}\mathcal{B}_\mathrm{rec}$) and check whether it ever becomes negative.  
Note that there is an up-canyon time-mean flow with a positive velocity shear, varying from $\sim$0\,m/s near the seafloor to $\sim$0.06\,m/s at 200\,m above the topography. This time-mean flow is likely a consequence of turbulent mixing rather than its cause \cite{garrett2001isopycnal}, and is negligible compared to the tidal velocities (Fig.~\ref{pnas_fig1}\textit{D}), so we exclude it from the calculations.

The kinematic model (Eq.~\ref{eq:reconstruct B}) reproduces the evolution of buoyancy, as demonstrated by the close alignment of color shading and black contours in Fig.~\ref{pnas_fig3}\textit{C}-\textit{D}.
However, when using the linear-fit velocity ($U_\mathrm{fit}$), the reconstructed vertical buoyancy gradient $\partial_{\tilde z} \mathcal B_\mathrm{rec}$ remains always positive (Fig.~\ref{pnas_fig3}\textit{D}), indicating that the background tidal shear is not strong enough to overturn the background stratification.

Alford et al.~\cite{alford2025buoyancy} and Naveira Garabato et al.~\cite{garabato2024convective} came to a different conclusion and argued that the observed shear is sufficient to overturn the background stratification and trigger convective instabilities. The discrepancy arises because they used the full velocity field, $u_\mathrm{obs}$, to advect the background stratification. We confirm in Fig.~\ref{pnas_fig3}\textit{C} that patches of unstable stratification do appear if one advects the background stratification with the full velocity field, $u_\mathrm{obs}$ (cyan contours). However, one is reversing cause and effect in this calculation. The patches of unstable stratification are generated by small-scale velocity fluctuations, which are the result but not the triggers of turbulent events \red{(Fig.~\ref{pnas_fig3}\textit{C-D}; shown later in Sec.~2)}.

It is worth making one final remark before leaving the convective instability hypothesis. In both Fig.~\ref{pnas_fig3}\textit{C} and \ref{pnas_fig3}\textit{D}, the initial stratification $\partial_{\tilde z} \mathcal{B}_\mathrm{rec}(t=0)$ is set to $N^2$, the time-mean value.
If the initial stratification was set to zero, \textit{i.e.}, $\partial_{\tilde z} \mathcal{B}_\mathrm{rec}(t=0)=0$, then $\partial_{\tilde z} \mathcal{B}_\mathrm{rec}$ would return to zero in subsequent tidal cycles, under advection by either velocity field, $u_\mathrm{obs}$ or $U_\mathrm{fit}$. However, this implicitly assumes that an overturn has already occurred at $t=0$ and does not explain what triggered it.

\vspace{10pt}
\noindent (2) \textbf{Test for the Kelvin--Helmholtz Instability Hypothesis}
\vspace{5pt}

\begin{figure}[t!]
\centering\includegraphics[width=0.9\linewidth]{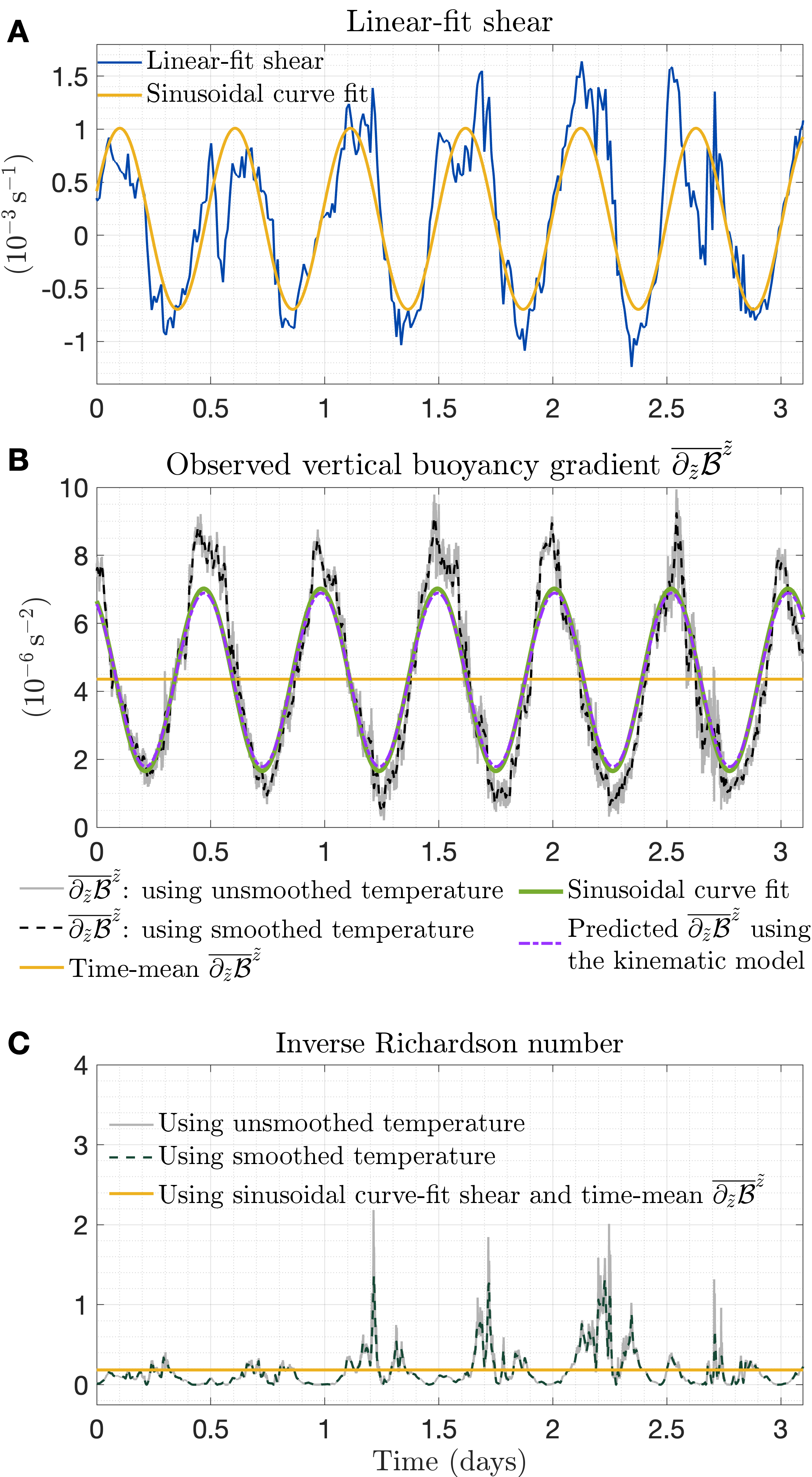}
\caption{
Time series from the MAVS2 mooring using the full depth measurements spanning from the seafloor to 224\,m above.
(\textit{A}) Slope-normal shear of the along-canyon tidal velocity estimated from a linear fit to velocity snapshots ($\Lambda_\mathrm{obs}(t)$, blue line), and a fit that is linear in the vertical and sinusoidal in time (yellow line).
(\textit{B}) Vertically averaged vertical buoyancy gradient $\overline{\partial_{\tilde z}\mathcal{B}}^{\tilde z}$. The gray line shows $\overline{\partial_{\tilde z}\mathcal{B}}^{\tilde z}$ computed using unsmoothed temperature data, with the linear relation between temperature and salinity (see Fig.~\ref{pnas_fig2}\textit{B}).
To reduce noise in the observed vertical buoyancy gradient $\partial_{\tilde z} \mathcal B$, a Gaussian filter with a 15-min window size is applied to the temperature field.
The black dashed line indicates $\overline{\partial_{\tilde z}\mathcal{B}}^{\tilde z}$ calculated from the smoothed temperature data.
The green line shows a sinusoidal curve fit.
The purple line represents predicted $\overline{\partial_{\tilde z}\mathcal{B}}^{\tilde z}$ based on the kinematic model (Eq.~\ref{predicted dBdz}). 
The yellow line represents the time-averaged and vertically averaged vertical buoyancy gradient.
(\textit{C}) Time series of the inverse Richardson number, calculated using the unsmoothed temperature (the gray solid line) and smoothed temperature (the dashed line). The yellow line represents \red{the inverse of the minimum $Ri$ of the background tide (Eq.~\ref{eq:Ri_min}).}
The $x-$axes indicate days since 7\,p.m. on 2021-07-07. 
}
\label{pnas_fig4}
\end{figure}

Throughout this paper, we refer to KH instability as the instability of a steady shear flow in a stratified fluid away from boundaries. KH instability arises when the Richardson number $Ri$ of the background shear drops below 1/4~\cite{drazin2002introduction}. Here, we wish to determine whether the BLT background tidal shear is KH unstable. 
\red{(Since most of this paper will later focus on dynamics that develop for Richardson numbers larger than 1/4, we will present the results in terms of the inverse Richardson number to map the interval from 1/4 to infinity to the interval between 0 and 4.)}
The first step, therefore, is to compute the linear shear associated with the tide within the canyon to distinguish it from the pointwise shear dominated by turbulent fluctuations. The blue curve in Fig.~\ref{pnas_fig4}\textit{A} shows the time series of the linear-fit shear $\Lambda_\mathrm{obs}$ from the MAVS2 mooring. The black dashed curve in Fig.~\ref{pnas_fig4}\textit{B} is the vertically averaged vertical buoyancy gradient $\overline{\partial_{\tilde z}\mathcal{B}}^{\tilde z}$. The Richardson number of the background flow, $Ri(t)=\overline{\partial_{\tilde z}\mathcal{B}}^{\tilde z}/(\Lambda_\mathrm{obs}\cos\theta)^2$, never drops below 1/4 as shown in Fig.~\ref{pnas_fig4}\textit{C}. \red{(Here, $\theta\approx5.6^\circ$ is the topographic slope at MAVS2 and $\cos\theta\approx0.995$.)} KH instability is unlikely to be the mechanism driving the onset of turbulence once per tidal period in the canyon. However, it is important to keep in mind that the $Ri<1/4$ criterion applies to stationary shear away from bottom topography. Both conditions are violated in the BLT canyon with important implications as we discuss in the following sections.

The analysis presented in Fig.~\ref{pnas_fig4} is based on a subset of 3 days of data, but Figs.~S1-S2 confirms that the results are representative of the whole 90-day observational period (Supporting Information). 

Further confirmation that the flow is stable to KH instability can be assessed with a simple theoretical model. The background tidal shear sweeping back and forth along the canyon can be described as a constant shear with a sinusoidal time dependence,
\begin{equation}
    u(z,t) = \Lambda z\cos(\omega t),
\end{equation}
where $\omega$ is the M2 tidal frequency, and $\Lambda$ is the magnitude of the velocity shear. The value of $\Lambda$ is obtained by fitting a sinusoidal curve to the observed linear-fit shear over time, shown as the yellow curve in Fig.~\ref{pnas_fig4}\textit{A}.
Using this, the kinematic model (Eq.~\ref{eq:reconstruct B}) predicts the vertically averaged vertical buoyancy gradient as 
\begin{equation}
    \overline{\partial_{\tilde z}\mathcal{B}}^{\tilde z}\big\vert_\mathrm{predict} = N^2 -N^2 \frac{\Lambda}{\omega}\sin(\omega t)\sin\theta\cos\theta.
    \label{predicted dBdz}
\end{equation}
In Fig.~\ref{pnas_fig4}\textit{B}, the purple curve shows the predicted vertical buoyancy gradient based on the equation above, while the green curve shows a sinusoidal fit to the observed vertical buoyancy gradient. The close alignment between the two curves further supports the reliability of the kinematic model. \red{The yellow line in} Fig.~\ref{pnas_fig4}\textit{C}, \red{computed using $\Lambda$ and $N^2$ (Eqs.~\ref{eq:Ri_full}--\ref{eq:Ri_min}),} further confirms that the Richardson number of the background tidal flow is substantially larger than one at all times.

\section{\red{Numerical and theoretical study of parametric shear instability in an abyssal canyon}}
\label{sec:gcm}

In the previous section, we demonstrated that the tidal flow in the BLT canyon is stable to convective and KH instability. We now examine its stability to parametric instability associated with the time-dependent shear using a high-resolution, non-hydrostatic, two-dimensional (2D) configuration of the Massachusetts Institute of Technology General Circulation Model (MITgcm) \citep{MarshallAdcroftHill97,MarshallHillPerelman97}.

\red{In the simulations, we impose a tidal force acting along the canyon axis. As shown in Fig.~\ref{pnas_fig1}\textit{D--E}, the canyon walls significantly suppress the impact of rotation on the background tide that would drive a cross-canyon tidal flow. Ma et al.~\cite{ma2025tide} shows that a cross-canyon tidal velocity can only develop for canyons wider than the local deformation radius, $L_R = NH/f\approx8\,$km, with $H\approx 500\,$m the depth of the canyon, $N\approx2\times10^{-3}\,\mathrm{s^{-1}}$ the typical background stratification, and $f\approx1.2\times10^{-4}\,\mathrm{s^{-1}}$ the Coriolis parameter. This is larger than the canyon width ranging from a few hundred meters to 5\,km. }

\red{While the large-scale tidal flow does not develop a cross-slope velocity, no such constraint applies to small-scale perturbations. However, BLT observations show that the turbulent overturns develop over the full 200~m tidal shear layer~\cite{garabato2024convective,alford2025buoyancy} and the entire canyon width~\cite{wynne_preprint}. We therefore choose to consider perturbations that span the full width of the submarine canyon where there is substantial tidal shear. In this thin-canyon approximation, the perturbations have no cross-canyon $v$-velocity to satisfy the no-normal flow boundary conditions at the lateral sidewalls. They are essentially Kelvin-waves modulated by the tidal shear often reported in canyons~\cite{baines1983tidal,grimshaw1985reflection}. We will show that these perturbations are parametrically unstable for the BLT flow parameters even though $Ri>1/4$. While additional perturbations with non-zero $v$ velocities---Poincar\'e waves modulated by the shear---could also be unstable, this will not affect the conclusion that parametric instability is a likely explanation for the observed turbulence in the BLT canyon; additional unstable waves can only contribute to the instability. On the other hand, there are advantages to making the thin-canyon approximation: the analysis of parametric instability is more transparent and the structure of the unstable models is not sensitive to variations of the canyon width with depth~\cite{ma2025tide}.}

\red{We will assess the impact of perturbations with a non-zero $v$ velocity at the end of the section by running numerical simulations of 3D tidal flows in a canyon. The additional modes result in similar growth rates, and the basic physics remains the same.}



\red{In the absence of any $v$-velocity, the dynamics is essentially two dimensional. We take advantage of this simplification to run high-resolution 2D simulations of the tidal flow instability.}
The model is configured in a $3,000\,\text{m}\times500\,$m domain, with additional tests using domain sizes up to $10\,\text{km}\times1500\,$m showing negligible impact on the results. Our focus is to study the growth of instability before the onset of turbulence. However, the model can resolve, at least qualitatively, also the turbulent mixing following the onset of instabilities thanks to grid spacings of 3\,m in the horizontal and 1\,m in the vertical, with time steps ranging from 0.5\,s to 10\,s depending on the tidal amplitude.

\begin{figure}[t!]
\centering\includegraphics[width=0.9\linewidth]{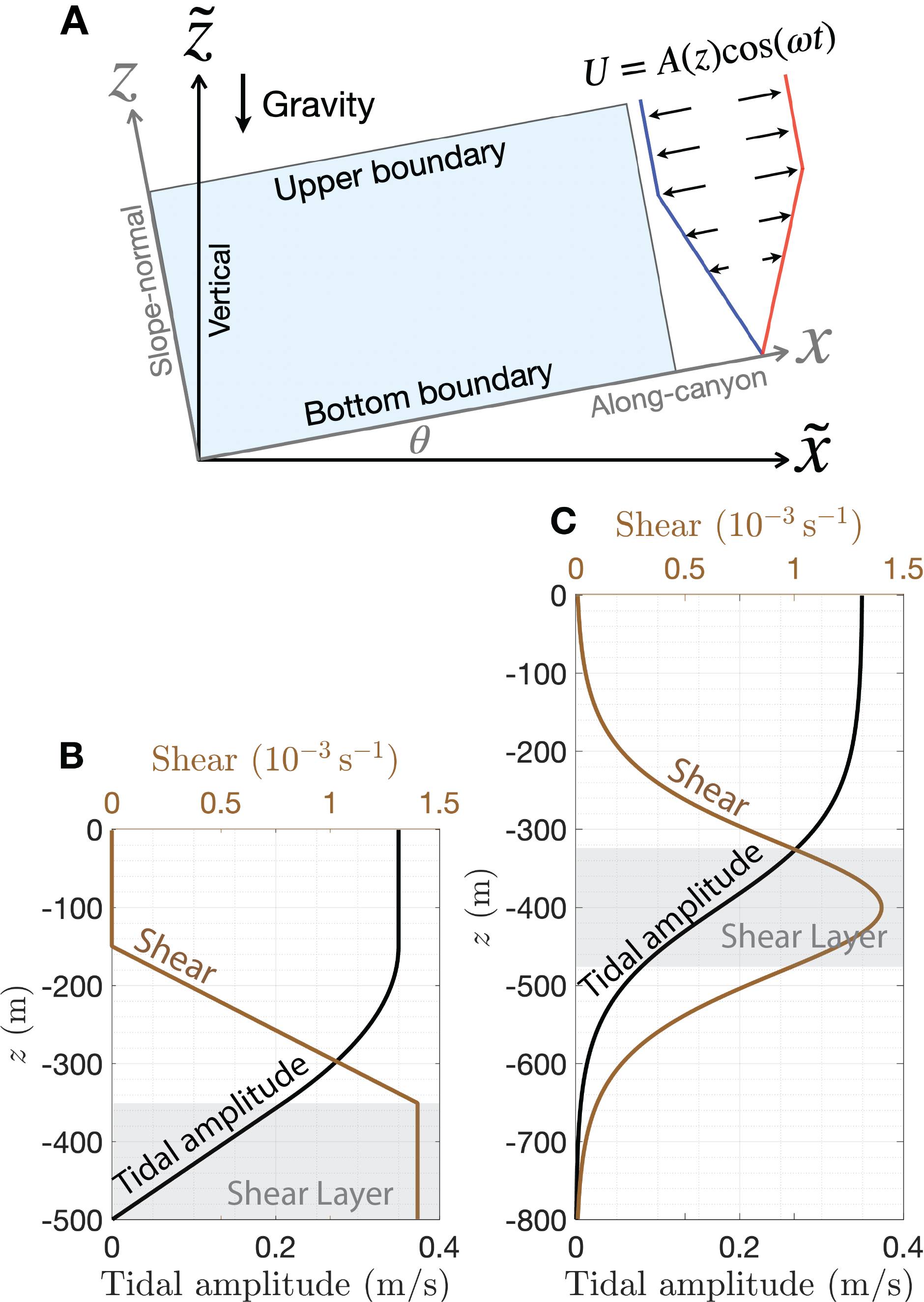}
\caption{
MITgcm configurations.
(\textit{A}) Schematic of the slope-aligned coordinate. The normal and slope-aligned coordinates are denoted by $(\tilde x,\tilde z)$ and $(x,z)$, respectively. 
(\textit{B}) Piecewise linear tidal current amplitude $A(z)$ (black, lower $x-$axis) and its shear $\Lambda=\partial_z A(z)$ (brown, upper $x-$axis). 
The tidal amplitude has been smoothed to avoid instabilities caused by any abrupt transition of shear.
The gray rectangle indicates the 150\,m shear layer used to compute the instability growth rate. Using a 250\,m shear layer yields the same growth rate.
(\textit{C}) Tidal current amplitude $A(z)$ as a hyperbolic tangent function (black, lower $x-$axis) and its shear $\Lambda=\partial_z A(z)$ (brown, upper $x-$axis). 
}
\label{pnas_fig5}
\end{figure}

We modify the non-hydrostatic Navier-Stokes equations solved by the MITgcm equations in two ways. First, we add a body force in the momentum equation to force a background tidal flow $U(z,t)$ of the form (see Methods),
\begin{equation}
    U(z,t) = A(z) \cos(\omega t) ,\ \mathrm{with}\  A(z) = \int_0^z \Lambda(z')\ \mathrm{d}z',
    \label{eq:U(z,t)}
\end{equation}
where $\omega$ is the tidal frequency, $\Lambda$ is the velocity shear and $z$ is the height above bottom. Two choices of $A(z)$ are considered: a linear function in $z$ and a hyperbolic tangent profile (Fig.~\ref{pnas_fig5}\textit{B}-\textit{C}). The resulting velocity field is then the sum of the forced tidal component, $U(z,t)$, and departures from it given by the dynamics, $\left(u(x,z,t),w(x,z,t)\right)$. Second, the total buoyancy $\mathcal B(x,z,t)$ is decomposed into a steady background component, $B_0(x,z)=N^2 (z\cos\theta + x\sin\theta)$, and a time-dependent component, $b(x,z,t)$. This is the same decomposition used in the observational analysis, but the time-dependent component is now function on both $x$ and $z$ because buoyancy is advected by both the background tidal shear, which depends on $z$ and $t$, and the velocity fluctuations $(u(x,z,t),w(x,z,t))$. The model solves for the evolution of $u(x,z,t)$, $w(x,z,t)$ and $b(x,z,t)$.

To explore the influence of the topographic slope on the parametric instability and subsequent onset of turbulence, we perform simulations over a sloping bottom inclined at the \red{mean} angle of the BLT canyon ($\theta=4^\circ$) and a flat bottom ($\theta=0^\circ$). For simulations with a sloping bottom, we rotate the buoyancy and momentum equations into the slope-aligned coordinate system (Fig.~\ref{pnas_fig5}\textit{A}; Methods) following previous studies \citep{drake2022dynamics,wenegrat2018submesoscale}. This approach simplifies the numerical representation of the bottom topography by preventing numerical instabilities associated with discrete steps in the natural coordinate system. In addition, it allows us to apply periodic boundary conditions for both velocities and the time-dependent buoyancy component in the along-canyon direction.

\begin{figure*}[!ht]
\centering
\includegraphics[width=0.83\linewidth]{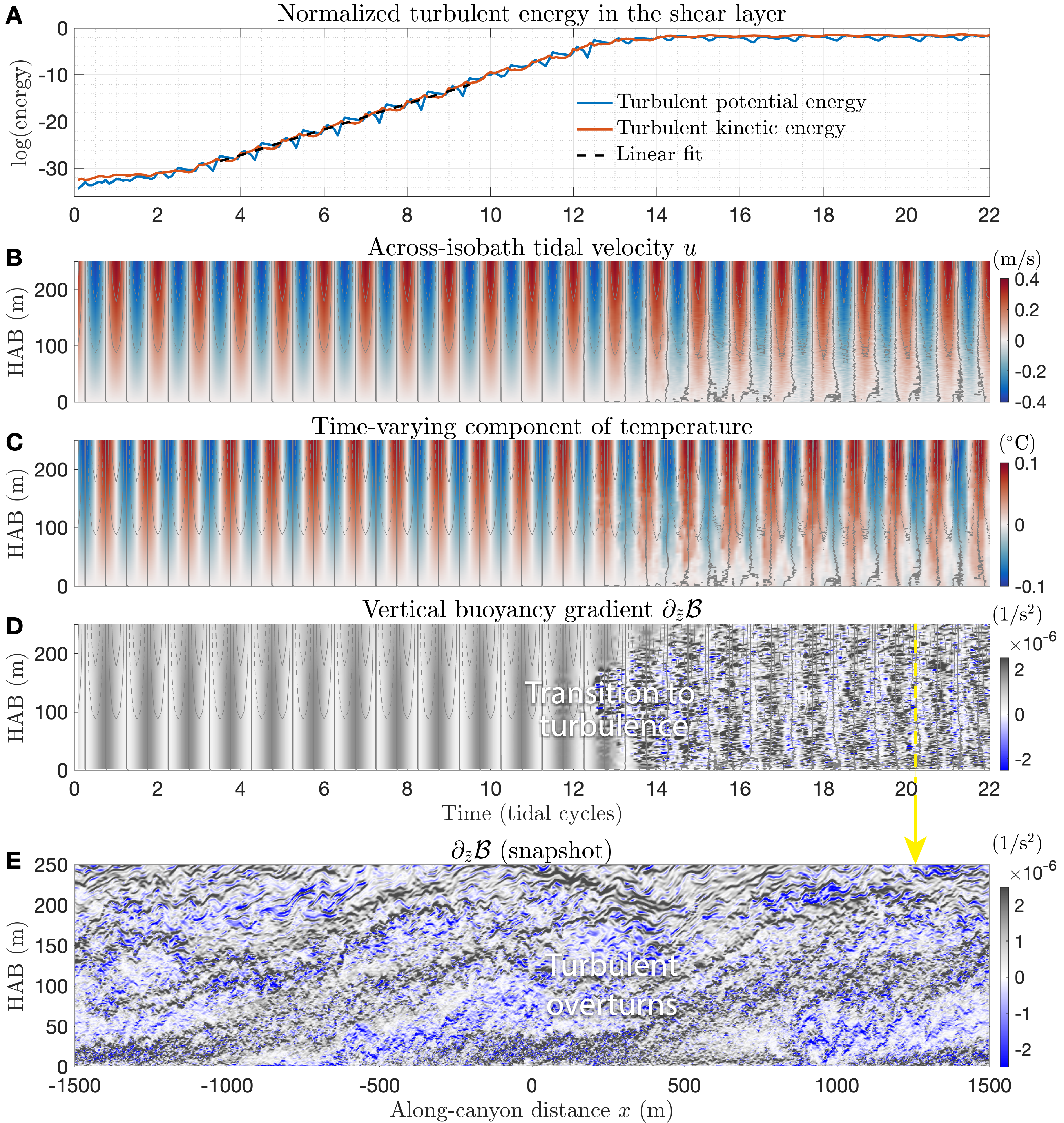}
\caption{
Results of the numerical simulation with a sloping topography ($\theta=4^\circ$) and piecewise linear tidal shear. \red{The minimum Richardson number of the background tide is $Ri_\mathrm{min} = 0.28$. 
Note that the results remain qualitatively similar for Richardson numbers of order unity; a small value of $Ri_\mathrm{min}$ is chosen here to better visualize the turbulent overturns.
}
(\textit{A}) Time series of the normalized turbulent potential energy and kinetic energy in the bottom shear layer as a function of time (tidal cycles). To determine the growth rate, a linear fit is applied to the logarithmic turbulent potential energy during the period of exponential growth (dashed black line).
(\textit{B}--\textit{D}) Time series of the along-canyon tidal velocity, time-varying component of potential temperature, and vertical buoyancy gradient ($\partial_{\tilde z}\mathcal{B}=\partial_z\mathcal{B}\cos\theta+\partial_x\mathcal{B}\sin\theta$) at a selected location ($x=300\,$m), as a function of time and the height above the bottom (HAB). Since the domain is periodic in the $x$-direction, the choice of location for the time series does not qualitatively affect the results. 
Thick gray contours denote zero along-canyon tidal velocity $u$, while thin solid and dashed gray contours represent positive and negative $u$ values, respectively, with an interval of $0.15$\,m/s. 
(\textit{E}) Snapshot of the vertical buoyancy gradient toward the end of the flood phase during the 20th tidal cycle (marked with a yellow arrow).
}
\label{pnas_fig6}
\end{figure*}

The simulations are initialized with infinitesimal white noise temperature perturbations on the order of $\sim10^{-20}\,^\circ\mathrm{C}$ to allow the growth of perturbations with any possible wavelength. The tidal forcing is chosen such that the $Ri$ of the resulting tidal flow is always larger than 1/4, ensuring that no KH instability can develop. Fig.~\ref{pnas_fig6} shows the temporal evolution of the turbulent kinetic energy (TKE) (Eq.~\ref{eq:MSDu} in Methods) and turbulent potential energy (TPE), $b^2/(2\red{N^2})$, for a simulation over a $4^\circ$ sloping topography. The TKE and TPE undergo periodic oscillations in response to the tidal forcing. But after an initial transient of \red{2$\sim$3} tidal cycles, the amplitudes of TKE and TPE at the end of each periodic oscillation is slightly larger than in the previous cycle, leading to an overall exponential increase in both TKE and TPE (Fig.\,\ref{pnas_fig6}\textit{A} and Fig.\,S3\textit{B}). This is a telltale signature of parametric instability.

Once the perturbations reach a finite amplitude, they lead to a burst of turbulence that overturns and mixes the density profile. This results in a sudden increase in mean potential energy---as mixing raises the center of mass of the fluid---and dissipation of TKE at molecular scales. This represents the transition from the linear growth rate phase of the instability to its nonlinear equilibration. The transition to finite amplitude perturbations takes approximately 10 \red{tidal cycles}, because we initialized the simulation with infinitesimal temperature noise to accurately compute the growth rate of the instability. In the real ocean, irregular topography, internal waves, and other flows will result in finite initial perturbations and a transition to turbulence within just a few tidal cycles. 
The turbulent phase lasts only a few hours, but then subsides as the tidal shear restratifies the fluid, only to be followed by another turbulent burst when the shear reverses sign and brings the density stratification back to zero (Figs.\,\ref{pnas_fig6}\textit{D},\,S3\textit{E}). This demonstrates that the parametric instability drives the first overturning event, but once that happens, no further instability is necessary because the tidal shear will bring the stratification back to zero every subsequent tidal cycle (see Eq.~\ref{eq:reconstruct B} setting $\mathcal{B}_\mathrm{rec}=0$), resulting in a regular sequence of overturning events as observed in the BLT field campaign.

\begin{figure*}[t!]
\centering\includegraphics[width=0.83\linewidth]{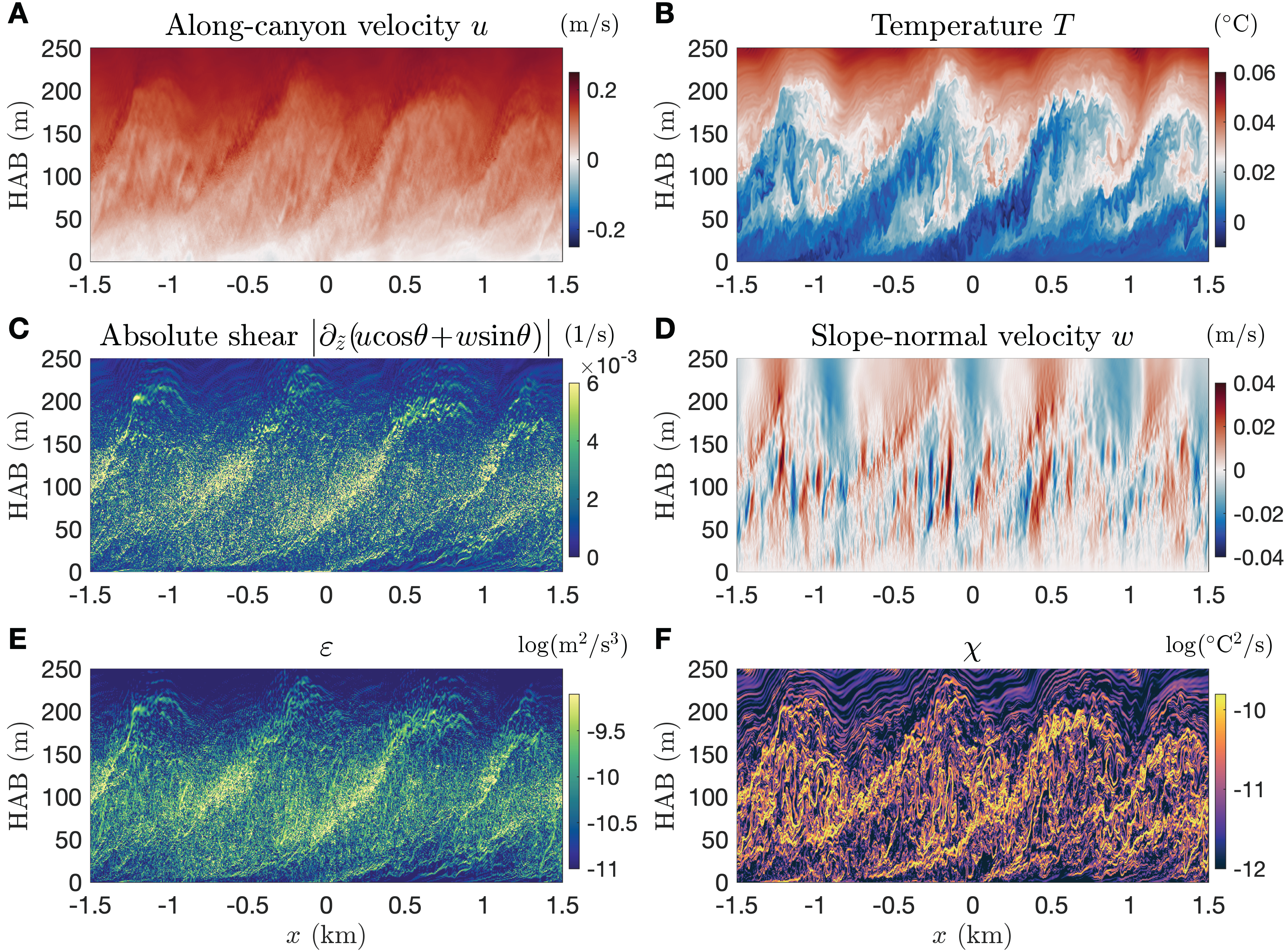}
\caption{
Snapshots of the MITgcm simulation with a sloping topography ($\theta=4^\circ$) and a piecewise linear tidal shear. The minimum Richardson number of the background tide is $Ri_\mathrm{min}= 0.28$. \red{Note that the results remain qualitatively similar for Richardson numbers of order unity; a small value of $Ri_\mathrm{min}$ is chosen here to better visualize the turbulent overturns.} The snapshots were captured toward the end of the flood phase during the 14th tidal cycle: (\textit{A}) along-canyon tidal velocity $u$, (\textit{B}) potential temperature $T$, (\textit{C}) absolute velocity shear $\vert\partial_{\tilde z} (u\cos\theta+w\sin\theta)\vert$, where $(u\cos\theta+w\sin\theta)$ is the horizontal velocity in the natural coordinate system, (\textit{D}) slope-normal velocity $w$, (\textit{E}) dissipation rate of turbulent kinetic energy $\varepsilon$, and (\textit{F}) destruction rate of temperature variance $\chi$ as a function of the height above the bottom (HAB) and the along-canyon distance $x$. Panels \textit{E} and \textit{F} are plotted on a logarithmic scale with a base of 10.}
\label{pnas_fig7}
\end{figure*}

Fig.~\ref{pnas_fig7} shows a snapshot during one of the turbulent mixing events. Large density overturns are evident in the temperature field\red{, $T$,} as well as in the along-canyon and slope-normal velocities (Fig.~\ref{pnas_fig7}\textit{A},\textit{B},\textit{D}). Additionally, regions of strong local velocity shear correspond to peaks in the dissipation rate of TKE\red{, $\varepsilon = 2 \nu_i \left( \frac{1}{2} ( \partial_j u_i + \partial_i u_j ) \right)^2$,} and the destruction rate of temperature variance\red{, $\chi = 2 \kappa_i (\partial_i T)^2$, where $i,j$ are summation indices, and $\nu$ and $\kappa$ are the constant viscosity and diffusivity, respectively} (Fig.~\ref{pnas_fig7}\textit{C},\textit{E},\textit{F}\red{; Table S1}). The vertical scale of the overturns spans the whole depth range of the background tidal shear, which is primarily controlled by the canyon's geometry~\cite{ma2025tide}.

In the Supporting Information, we show that the growth of the parametric instability is essentially the same over the 4 degree sloping topography as over a flat bottom (Fig.~S3). However, the topographic slope affects the timing of the turbulent bursts. Over sloping topography, turbulent overturns develop during the flood tidal phase, when the tidal shear reduces background stratification. During the ebb phase, the negative shear re-stratifies the flow, thereby suppressing mixing. This is consistent with the observed onset of turbulence during positive tidal shear in the BLT canyon (Fig.\,\ref{pnas_fig2}\textit{A},\textit{C}) \citep{van2024near,garabato2024convective,wynne2024observations}. Over a flat bottom, the tidal velocity is parallel to the background density surfaces and does not affect the stratification of the background field. The simulation shows that turbulence develops twice per tidal cycle, when the \red{magnitude of the} tidal shear is largest (Fig.~S3\textit{E}).

Having established that the tidal flow is unstable in a setup representative of the BLT environment, we now focus on analyzing the nature of the instability. Both the shape of the background velocity profile and its time dependence influence the stability of the sheared flow.
For a steady shear flow: 
(i) If the velocity shear is \textit{constant} in the vertical direction, perturbations exhibit a transient growth before eventually decaying, regardless of the shear magnitude \cite{SiFerrari25b,shepherd1985time}; 
(ii) If the velocity profile has an \textit{inflection point}, where its curvature changes sign, KH instability can arise when the minimum Richardson number falls below 1/4~\cite{miles1961stability,howard1961note};
(iii) If the velocity shear is in contact with a \textit{solid boundary}, KH instability is suppressed~\cite{liu2023effects}. None of these scenarios is relevant to our simulations because the Richardson number is always larger than 1/4. On the other hand, introducing time dependence into the background shear has been shown to destabilize the flow compared to the steady problem~\cite{kelly1965stability,radko2019instabilities} and we therefore focus on this aspect.

\begin{figure}[t!]
\centering
\includegraphics[width=0.9\linewidth]{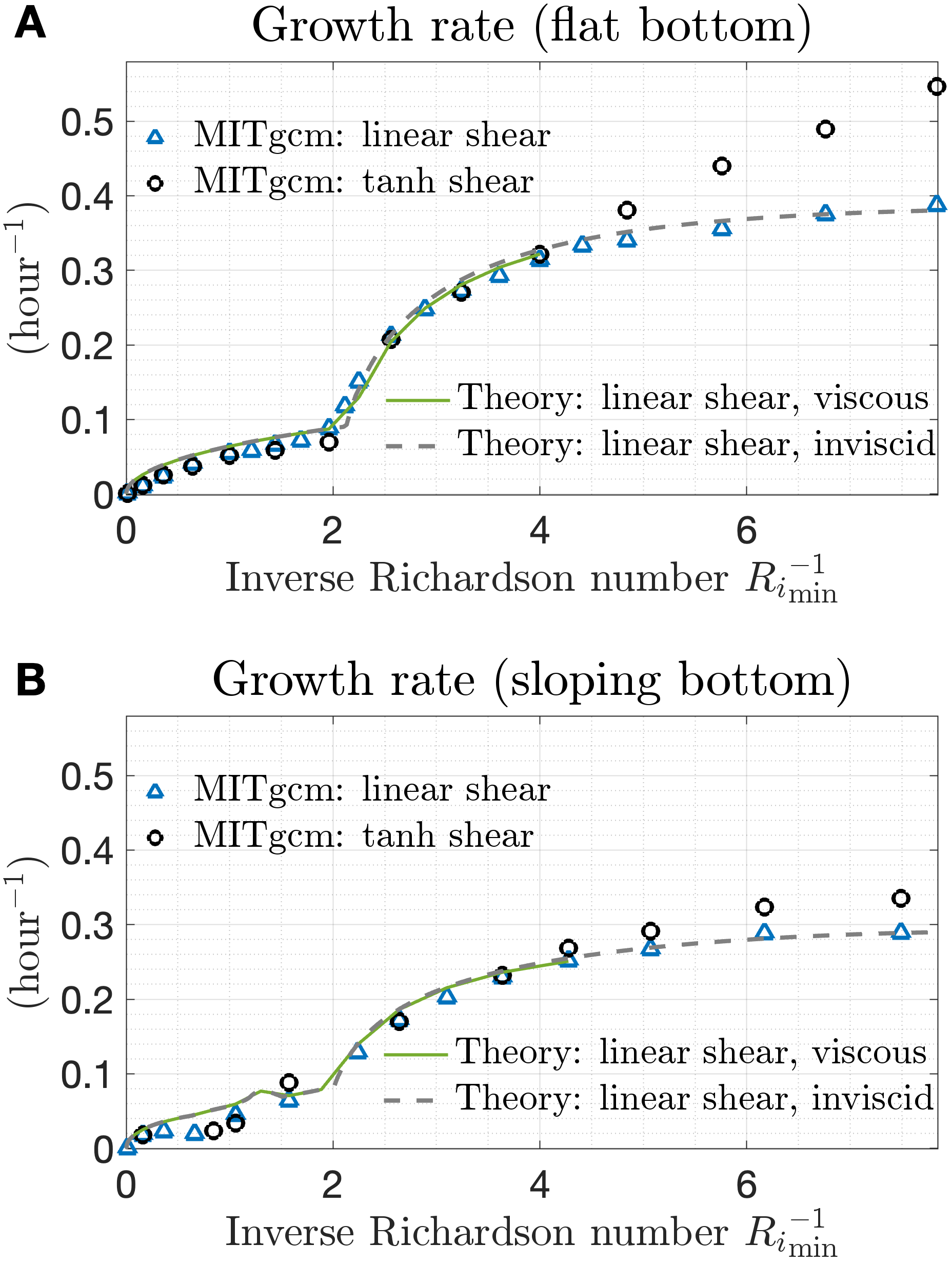}
\caption{
Instability growth rate as a function of $Ri_\mathrm{min}^{-1}$ for (\textit{A}) flat bottom and (\textit{B}) sloping bottom.  
The blue and black markers represent the MITgcm simulations with a piecewise linear tidal amplitude profile (Fig.~\ref{pnas_fig5}\textit{B}) and a hyperbolic tangent profile (Fig.~\ref{pnas_fig5}\textit{C}), respectively.
The green and gray lines represent the growth rates of the most unstable modes in the viscous theory and inviscid theory, respectively (Si \& Ferrari, 2025b; ref.~\citealp[]{SiFerrari25b}).}
\label{pnas_fig8}
\end{figure}

Fig.~\ref{pnas_fig8} shows the growth rate of the instability for a series of simulations where we vary the magnitude of the background tidal shear, but keep the same oscillation frequency and background stratification.
The growth rates are reported as a function of the smallest Richardson number reached by the background flow during a tidal period (Eqs.~\ref{eq:Ri_full}--\ref{eq:Ri_min} in Methods).
The oscillatory tidal shear is unstable for all $Ri$, including values above 1/4 (blue markers).
The black lines show the theoretical growth rate computed using Floquet theory, designed to study the instability of oscillatory flows \cite{SiFerrari25b}.
While the technical details are tedious and numerically delicate, the basic idea is straightforward. 
One considers substitutes perturbations of the form $(u,w,b)=(u_0(z),w_0(z),b_0(z))e^{ikx-i\omega t+\sigma t}$ in the governing equations (Eq.~\ref{eq:mitgcm} in Methods, where $k$ is the horizontal wavenumber of the perturbed waves, and ($u_0, w_0, b_0$) are the perturbation amplitudes), discards nonlinear terms, and computes the growth rate $\sigma$ after one tidal period.
The calculation details are reported in a companion paper (Si \& Ferrari, 2025b; \citealp[]{SiFerrari25b}).
The theoretical growth rate matches the estimate from the numerical simulations providing compelling support for the hypothesis that the instability is indeed associated with the time dependence of the tidal shear. 
Indeed Si \& Ferrari (2025b) show that the equations for the perturbations can be reduced to the standard equation for parametric instability, \textit{i.e.}, the equation of a harmonic oscillator with a frequency that changes periodically in time \cite{SiFerrari25b}.

To assess whether the onset of the instability and the growth rates in the oscillatory shear flows are influenced by the vertical profile of the background tides, we additionally conduct MITgcm simulations using hyperbolic tangent velocity profiles (Fig.~\ref{pnas_fig5}\textit{C}), a profile that allows KH instability to develop when $Ri_\mathrm{min} < 1/4$. 
Fig.~\ref{pnas_fig8} demonstrates that, when $Ri_\mathrm{min}>1/4$, the growth rate of oscillatory shear instability is mostly unaffected by the structure of the background tides. 
However, when $Ri_\mathrm{min} < 1/4$, the flow becomes more unstable under an oscillating hyperbolic tangent velocity profile than under an oscillating linear velocity profile, due to the combined effects of KH instability and parametric instability (Fig.~\ref{pnas_fig8}).

\red{The instability analysis focused on perturbations without any $v$ component. To address the impact of $v\neq 0$ perturbations, we ran 3D MITgcm simulations in a 2.4~km-wide canyon with no restrictions on the cross-canyon velocities except that they must vanish at vertical lateral sidewalls (Methods and Supporting Information). The presence of lateral sidewalls results in noisier simulations due to the lateral reflection of waves and perturbations, but the solutions do undergo parametric instability with growth rates within 40\% of those based on $v=0$ perturbations. This confirms that our results are qualitatively and quantitatively robust despite the thin-canyon approximation for the perturbations. A detailed analysis of parametric instability of sheared flows with and without rotation in the open ocean and canyons is the focus of a companion paper~\cite{SiFerrari25b}. 
}

\section{Conclusions and Discussion}
\label{sec:conclusion}

We investigated the physics driving the turbulent overturns observed during the BLT field campaign. 
Our analysis showed that the large-scale tidal shear was too weak to drive convective instability by advecting dense water over lighter water, or to drive steady-shear KH instability.
Using idealized MITgcm simulations with prescribed tidal shear, we identified parametric instability, which is associated with the oscillatory nature of the tidal flow, as the primary mechanism driving turbulent overturns.
Consistent with previous studies, the oscillatory shear flow is unstable even for Richardson numbers of order 1, which would typically indicate stable conditions \red{for steady-shear KH instability}.
The oscillatory instability is crucial for triggering the first onset of turbulence. After the first turbulent overturning event starts, turbulent mixing significantly reduces the vertical buoyancy gradient, enabling turbulent overturns to recur in subsequent tidal cycles.

We found that the parametric instability is not affected much by the sloping topography, with growth rates similar to those found over a flat bottom. However, the presence of a slope determines the timing of turbulent events; as revealed by both the BLT observations and MITgcm simulations, overturning events occur when the tidal shear is positive over the slope, as a result of \red{the reduction of} stratification associated with differential advection during this phase. 

\red{We focused our instability analysis on 2D time-dependent shears to simplify the analysis and interpretation of the results. However, we also ran 3D simulations and found similar results, further supporting our conclusion that time-dependent shears are unstable at Richardson numbers larger than one quarter.}

\red{These results emphasize that tidal shears are more unstable than steady shears and can trigger turbulent overturns and mixing in canyons. There is a large body of both theoretical and observational literature showing that tidal shears are common in canyon geometries~\citep[\textit{e.g.}][]{hotchkiss1982internal,baines1983tidal,grimshaw1985reflection,hall2011internal,zhang2014modeling,aslam2018internal,masunaga2023numerical}. Ma et al.~\citep{ma2025tide} recently demonstrated that steep canyons, such as those examined during the BLT campaign, can trap standing internal waves oscillating back and forth along the canyon axis. The waves are associated with substantial shears on a vertical scale of the order of the canyon depth. Our finding that such time-dependent shears are parametrically unstable suggests that deep canyons are likely significant hotspots for deep ocean mixing. Instead the continental shelves outside the canyons may be less suited to support shear instabilities because the canyon geometry with lateral walls is key to generate strong tidal shears--except for the special case when the shelf slope is critical to internal tides. (A separate discussion pertains to latitudes where the Coriolis frequency matches the tidal one, 72$^\circ$N for M2 and 29$^\circ$N for K1~\cite{mackinnon2013parametric,ansong2018geographical}. At these latitudes resonances are likely to enhance parametric instabilities both in canyons and the surrounding shelves.)}

\red{ 
Our analysis suggests that parametric instability is a likely driver for the turbulent overturns observed once per tidal cycle in the BLT canyon. We are not arguing that other instabilities cannot be at play. First, the enhancement of shears near the seafloor through bottom stresses can trigger traditional KH instabilities trapped a few meters to the bottom. Second, horizontal shears along the steep canyon walls can produce pancake-shaped vortices and turbulent patches~ \cite{cope2020dynamics,garaud2024combined}. Third, finite amplitude perturbations due to irregular topography or small-scale internal waves could also push the system over the instability threshold. However, all these other instability pathways depart from the parametric one in that they would result in localized turbulent patches rather than the large vertical overturns observed in BLT Canyon~\cite{wynne2024observations}.}

\red{We are not arguing that turbulence in deep ocean canyons is only associated with parametric instabilities of tidal shears, but rather that tidal shears are much more likely to support turbulence in deep ocean canyons because parametric shear instabilities arise for much weaker shears than steady shear instabilities. The geometry, stratification, and tidal dynamics of the BLT canyon are representative of many other abyssal canyons \cite{kunze2012turbulent,van2022contrasting,alberty2017reflecting,wynne2024observations}, and the instability we described is quite generic and depends only weakly on the slope or width of the canyons. Since the upwelling of the deep branch of the global MOC relies on abyssal mixing to convert dense to light waters, our results suggest that parametric instability of tidal flows may play an important role in driving the abyssal ocean circulation.
}


\red{More broadly, this} study suggests that accurately representing turbulence induced by oscillatory flows could be important for other systems, such as (i) tide-induced melting of sloping ice shelves, (ii) paleoclimate periods with stronger tides than present, as well as (iii) ice-covered moons with strong tidal flows in their liquid oceans \citep{tyler2008strong}.


\matmethods{
\subsection{\red{Two-dimensional model configuration}}
The MITgcm simulations are implemented with the 7th-order one-step advection scheme, a monotonicity-preserving limiter, and a quadratic bottom-drag coefficient of $2.5\times10^{-3}$. 
To minimize numerical errors, we apply an isotropic eddy diffusivity ($\kappa$) and eddy viscosity ($\nu$) of $5\times10^{-6}\,$m$^2$/s for the flat-bottom simulations, and $1\times10^{-5}\,$m$^2$/s for the sloping-bottom simulations. 
These values were selected to ensure numerical stability while keeping viscosity and diffusivity as low as possible.
For simplicity, we use a linear equation of state that depends solely on temperature: $\rho = \rho_0\big(1-\alpha(T-T_0)\big)$, where $\rho_0=999.8\,$kg/m$^3$ is the reference density, $T_0=0^\circ$C is the reference temperature, and $\alpha=2\times10^{-4} \, ^\circ$C$^{-1}$ is the constant thermal expansion coefficient. 

The total buoyancy, $\mathcal{B}(x,z,t)=-g(\rho-\rho_0)/\rho_0 = g\alpha T$, is decomposed into two components: a steady background buoyancy $B_0(x,z)$ and a time-varying component $b(x,z,t)$, which includes both the perturbations generated by tidal advection and turbulent flows. The model simulates the time-varying buoyancy component $b(x,z,t)$, where
\begin{equation}
   \mathcal B(x,z,t) \!=\! B_0(x,z)\!+\!b(x,z,t) \!=\! B_0(x,z)\!+\!B(z,t)\!+\!b'(x,z,t).
\label{eq:buoyancy decomposition}  
\end{equation}
Here, the time-dependent component $b(x,z,t)$ is further decomposed into buoyancy perturbations $b'(x,z,t)$ and a time-varying component due to large-scale tidal advection $B(z,t)$. The component $B(z,t)$ satisfies the following equations:
\begin{equation}
    \partial_t B+ U\partial_x B_0=0,\ \text{and} \ \overline{B}^t = 0,
    \label{eq:B}
\end{equation}
where $\overline{\ \bullet\ }^t$ denotes the time average over multiple tidal cycles. By substituting $U(z,t)=A(z)\cos(\omega t)$ and $B_0 = N^2(z\cos\theta+x\sin\theta)$ into Eq.~\ref{eq:B}, we obtain the expression for $B$:
\begin{equation}
    B(z,t) = -A(z)\omega^{-1}N^2 \sin\theta\sin(\omega t).
    \label{eq:B(z,t)}
\end{equation}
Similarly, we can split the total pressure field into a steady background pressure $P_0(x,z)$ in hydrostatic equilibrium and a time-varying pressure field $p(x,z,t)$.

Below we show the rotating equations for the time-varying component of buoyancy $b$, along-canyon velocity $u$, slope-normal velocity $w$, and continuity in the slope-aligned coordinate \citep{drake2022dynamics,wenegrat2018submesoscale}, with underlines indicating modifications to the original MITgcm equations. \red{We set $v=0$ to satisfy the no-normal flow boundary conditions at the lateral sidewalls, appropriate for waves whose lateral scale exceeds the canyon width. The model thus simulates the} velocity field $\mathbf u(x,z,t)=u\mathbf{\hat i}+w\mathbf{\hat k}$, but only the time-dependent components of the buoyancy and pressure fields.
\begin{equation}
\left\{
\begin{aligned}
&\partial_t u + \mathbf  u\cdot \nabla u = - \partial_x p + \underline{F_\mathrm{tide}(z,t)} +b\,\underline{\sin\theta} +\nabla\cdot(\nu \nabla u)  \\
& \red{f u = -\partial_y p}\\
&\partial_t w + \mathbf  u\cdot \nabla w = -\partial_z p +b\,\underline{\cos\theta} + \nabla\cdot(\nu \nabla w) \\
&\partial_t b + \mathbf u\cdot \nabla b + \underline{wN^2\cos\theta + uN^2\sin\theta}= \nabla\cdot(\kappa \nabla b) \\
&\nabla\cdot\mathbf u =0
\end{aligned}
\right.
\label{eq:mitgcm}
\end{equation}
\red{In this thin-canyon approximation, the cross-canyon momentum equation becomes a diagnostic equation for the cross-canyon pressure gradient $\partial_y p$ that balances the $u$ velocity.}

Here, $F_\mathrm{tide}$ is an idealized body force representing the semidiurnal M2 tide, with a period of 43,200$\,$s. 
We assume that the tide-generating force $F_\mathrm{tide}$ acts solely in the along-canyon direction ($x-$direction), described by the following equation:
\begin{equation}
    \partial_t U = F_\mathrm{tide}+B\sin\theta.
    \label{eq:Ut}
\end{equation}
This equation states that the tidal velocity tendency results from the tide-generating force and the tidally induced buoyancy change mapped to the slope-aligned coordinate.
Using Eq.~\ref{eq:B(z,t)}, we can derive the tide-generating force as
\begin{equation}
    \!\!\!\! F_\mathrm{tide}(z,t) = \partial_t U -B\sin\theta =A(z)\big(N^2\sin^2\theta-\omega^2\big)\omega^{-1}\sin(\omega t).
\end{equation}
By incorporating $F_\mathrm{tide}$ in the along-canyon momentum equation (Eq.~\ref{eq:mitgcm}), we impose the background flow in the desired form (Eq.~\ref{eq:U(z,t)}).

\subsection{\red{Instability growth rate}}
To quantify the instability growth rate, we compute the mean square deviation of velocity ($\mathrm{MSD}_{\mathbf u}$) at each $z-$level in the shear layer (gray rectangles in Fig.~\ref{pnas_fig5}) with respect to its hourly- and along-canyon-mean.
\begin{equation}
    \mathrm{MSD}_{\mathbf u} = \frac{1}{L}\int_{x=0}^{x=L}\bigg(\overline{\mathbf u}^{t=1\,\mathrm{hour}} - \overline{ \overline{\mathbf u}^{t=1\,\mathrm{hour}}}^{x} \bigg)^2 \mathrm{d}x,
    \label{eq:MSDu}
\end{equation}
where $L$ is the domain width in the $x-$direction and overlines indicate averages.  
The turbulent kinetic energy is defined as half of $\mathrm{MSD}_{\mathbf u}$ integrated in the shear layer. Similarly, we can define the turbulent potential energy using the mean square deviation of buoyancy.  
We compute the instability growth rate for MITgcm simulations using the best linear fit to \red{$\log_{10}$ of} the normalized turbulent energy in the shear layer (Fig.~\ref{pnas_fig6}\textit{A}).

\subsection{\red{Background Richardson number}}
The background Richardson number is a function of time, defined as
\begin{equation}
  Ri(t)  = \frac{\partial_{\tilde z} B_0+\partial_{\tilde z} B}{(\partial_{\tilde z} U)^2} = \frac{N^2 - N^2\sin\theta\cos\theta\Lambda\omega^{-1}\sin(\omega t)}{\big(\cos\theta\Lambda\cos(\omega t)\big)^2}.
  \label{eq:Ri_full}
\end{equation}
The minimum Richardson number of the background tide is 
\begin{equation}
    \!\!\!\!\!Ri_\mathrm{min} \!=\! \Big(\max\big(Ri^{-1} (t)\big)\Big)^{-1} \red{\!=\! \frac{1}{2} \frac{N^2}{\Lambda_c^2\cos^2\theta} \frac{1}{1-\sqrt{1-\Lambda^2/\Lambda_c^2}},} 
    \label{eq:Ri_min}
\end{equation}
\red{where $\Lambda_c = \omega/ (\sin\theta\cos\theta)$ is the critical shear for convective instability; \textit{i.e.}, the background vertical buoyancy gradient ($\partial_{\tilde z}B_0+\partial_{\tilde z}B$) reaches zero when the velocity shear equals $\Lambda_c$.}

\subsection{\red{Three-dimensional simulations}}

\red{
To assess the influence of rotation and non-zero cross-canyon velocity on the parametric instability and ensuing turbulence, we additionally conduct 3D non-hydrostatic MITgcm simulations, incorporating the Coriolis terms.
}

\red{
The 3D model is configured in a domain of $7.7\,\text{km (along-canyon, $x$)}\times 2.4\,\text{km (cross-canyon, $y$)}\times500\,$m (slope-normal, $z$). The grid spacings are 10\,m in both the $x$ and $y$ directions, and 2\,m in the $z$ direction. To ensure numerical stability, the horizontal (vertical) diffusivity and viscosity are set to $2\times10^{-5}$ ($4\times10^{-5}$)$\,\mathrm{m^2/s}$.}

\red{
Similar to the 2D configuration (Eq.~\ref{eq:mitgcm}), the governing equations are rotated from the standard vertical--horizontal coordinate system to the slope-aligned coordinate system. The model solves for the evolution of momentum and buoyancy perturbations in a stably stratified fluid forced by a tidal potential. Following the same approach as in Eq.~\ref{eq:mitgcm}, we only solve for the perturbations to allow for periodic boundary conditions in the along-canyon ($x$) direction. The buoyancy equation is the same as in Eq.~\ref{eq:mitgcm}, and the 3D momentum equations in this rotated framework are
\begin{equation}
\begin{aligned}
&\partial_t u + \mathbf  u\cdot \nabla u\, \underline{-fv\cos\theta} = - \partial_x p + \underline{F_\mathrm{tide}(z,t)} +b\,\underline{\sin\theta} +\nabla\cdot(\nu \nabla u)  \\
&\partial_t v+ \mathbf  u\cdot \nabla v \, \underline{+fu\cos\theta+fw\sin\theta}= -\partial_y p+ \underline{F_\mathrm{PGF}(z,t)}+\nabla\cdot(\nu \nabla v) \\
&\partial_t w + \mathbf  u\cdot \nabla w \, \underline{+fv\sin\theta}= -\partial_z p +b\,\underline{\cos\theta} + \nabla\cdot(\nu \nabla w) \\
\end{aligned}
\label{eq:mitgcm_3D}
\end{equation}
The underlines indicate deviations from the standard MITgcm formulation of the momentum and buoyancy budgets due to the rotated reference system and the imposed background stratification.
}

\red{
We conducted 3D simulations with lateral walls in the $y-$direction and periodic boundary conditions in the $x-$direction. 
A piecewise linear background tidal current, $U(z, t) = A(z)\cos(\omega t)$, is imposed in the along-canyon direction, with its amplitude $A(z)$ illustrated in Fig.~\ref{pnas_fig5}\textit{B}. In the narrow BLT canyon, the tidal flow exhibits rectilinear polarization (Fig.~\ref{pnas_fig1}\textit{D}-\textit{E}) because a pressure gradient force develops in the cross-canyon direction to satisfy the no-normal flow conditions at the lateral walls. To replicate this effect, we include a time and depth-dependent pressure gradient term in the cross-canyon direction, $F_\mathrm{PGF}(z,t)=fU(z,t) = fA(z) \cos(\omega t)$, which eliminates the background tides in the cross-canyon direction. Therefore, the Coriolis force acts only on the perturbations, while the background tides remain rectilinear.
}

\red{The 3D simulations are qualitatively similar to the 2D ones (Figs.~S5--S6, Supporting Information). The wave reflections between the walls cause fluctuations that lead to a rapid initial increase in TKE and TPE. The turbulent overturns are less pronounced in the 3D simulations than in the 2D ones, due to the coarser horizontal and vertical resolution, as well as the larger viscosity and diffusivity used to ensure numerical stability. Despite these differences, the instability growth rates in the 3D simulations with lateral walls are comparable to the 2D ones (Supporting Information).}



}
\showmatmethods{} 

\subsection*{Data and Code Availability} 
\small
\red{All code, data used to generate the figures, and experimental configurations from this study are available at: \href{https://doi.org/10.5281/zenodo.16938421}{https://doi.org/10.5281/zenodo.16938421}.}
The moored observations \citep{voet24moored} are available at \href{https://doi.org/10.5061/dryad.v15dv424f}{https://doi.org/10.5061/dryad.v15dv424f}, and the bathymetry dataset \citep{voet24bathymetry} can be found at \href{https://doi.org/10.17882/99872}{https://doi.org/10.17882/99872}.

\acknow{The authors are supported by the NSF OCE-1756324 \red{(R.F.) and OCE-1756264 (G.V.)}. We thank the MITgcm team for making their code available. 
\red{We acknowledge high-performance computing support from the Derecho system (doi:10.5065/qx9a-pg09) provided by the NSF National Center for Atmospheric Research (NCAR).}
We thank the BLT (Bottom Boundary Layer Turbulence) team, especially Alberto Naveira Garabato, Matthew Alford, Bethan Wynne-Cattanach, and Carl Spingys, for sharing their observational data and for useful discussions. We sincerely appreciate useful discussions on this work with Yuchen Ma, Keaton Burns, Glenn Flierl, Kurt Polzin, Andre Souza, Henri Drake, and Xiaozhou Ruan.
\red{We thank the editor and the two anonymous reviewers for their insightful comments.}
}
\showacknow{} 

\bibsplit[32]
\bibliography{main.bib}



\section*{Supplementary material}

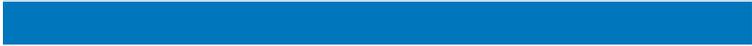

# Supporting Information for

**Tidally Induced Turbulence in the Abyssal Ocean**

**Yidongfang Si, Raffaele Ferrari and Gunnar Voet**

**Corresponding Author: Yidongfang Si.**
**E-mail: y_si@mit.edu**

**This PDF file includes:**

  Supporting text
  Figs. S1 to S7
  Table S1
  SI References

## Supporting Information Text

### 1. Observation

The yellow lines in Fig. S1 show the time series of stratification, $N^2$, at the MAVS moorings, calculated by applying a moving average with a 4-M2-tidal-cycle sliding window (equivalent to 2 days and 100 minutes) to the time series of the depth-averaged vertical buoyancy gradient. The low-passed value of $N^2$ changes little over time, with regions closer to the bottom exhibiting a weaker stratification. For example, at MAVS2, $N^2$ ranges from $3 \times 10^{-6}$ to $5 \times 10^{-6}\,\text{s}^{-2}$ for the bottom 224 m, and from $3 \times 10^{-6}$ to $4 \times 10^{-6}\,\text{s}^{-2}$ for the bottom 96 m. Additionally, Fig. S2 shows that the magnitude of the linear-fit shear at the MAVS moorings is broadly consistent with that in Fig. 4*A* in the main manuscript, despite variations associated with spring-neap tidal cycles. Overall, the Richardson number of the background flow remains close to one over long timescales.

### 2. Two-dimensional (2D) MITgcm simulations

Fig. S3 shows the results of the 2D MITgcm simulation over a flat bottom. Fig. S4 shows the time series of the 2D MITgcm simulation over a 4° sloping bottom.

### 3. Three-dimensional (3D) MITgcm simulations

Figs. S5–S6 present results from the 3D MITgcm simulations over a 4° sloping bottom in a channel laterally bounded by vertical walls, for $Ri_{\min} = 0.28$ (Fig. S5) and $Ri_{\min} = 1$ (Fig. S6), respectively (see Section D in Methods). The parametric instability growth rates for the 2D (i.e., assuming $v = 0$) and 3D simulations over a 4° sloping bottom are respectively:

| $Ri_{\min}$ | 2D | 3D |
|---|---|---|
| 0.28 | 0.23 hour$^{-1}$ | 0.13 hour$^{-1}$ |
| 1 | 0.05 hour$^{-1}$ | 0.04 hour$^{-1}$ |

The 3D channel with lateral walls exhibits slower growth rates than the 2D cases, likely due to the lower resolution and the higher diffusivity and viscosity employed in the 3D model.

### 4. Bulk mixing coefficient

We calculated the time series of the bulk mixing coefficient, $\Gamma_{bk}$, for the 3D simulations, which is defined as

$$\Gamma_{bk} = \frac{\alpha g}{2\langle \partial_{\bar{z}} T \rangle} \frac{\langle \chi \rangle}{\langle \varepsilon \rangle}. \qquad [1]$$

Here, $\alpha = 2 \times 10^{-4}\,°\text{C}^{-1}$ is the thermal expansion coefficient, $g = 9.8\,\text{m}^2/\text{s}$ is the gravitational acceleration, $\partial_{\bar{z}} T$ is the vertical temperature gradient, $\chi = 2\kappa_i (\partial_i T)^2$ is the destruction rate of temperature variance, and $\varepsilon = 2\nu_i \left(\frac{1}{2}(\partial_j u_i + \partial_i u_j)\right)^2$ is the dissipation rate of turbulent kinetic energy. The angle brackets $\langle \cdot \rangle$ denote volume averages computed over all horizontal grid points and vertically over the shear layer (bottom 150 meters).

Fig. S7 presents the time series of $\langle \partial_{\bar{z}} T \rangle$, $\langle \chi \rangle$, $\langle \varepsilon \rangle$, the ratio $\langle \chi \rangle / \langle \varepsilon \rangle$, and $\Gamma_{bk}$ for the 3D simulation with a 4° sloping bottom and $Ri_{\min} = 1$, corresponding to the simulation in Fig. S6. During the 10th tidal period, which marks the onset of turbulence (Fig. S6*D*), $\Gamma_{bk}$ ranges from about 0.15 to 0.7 (Fig. S7*E*). For comparison, the mixing coefficient observed in the BLT canyon is approximately 0.2 near the canyon top and increases to 0.3–0.7 toward the bottom (1). Once turbulence develops, however, $\Gamma_{bk}$ decreases rapidly (Fig. S7*E*) because stratification is not maintained in this idealized setup (unlike in the real ocean) and the coarse 3D resolution limits the model's ability to fully resolve turbulence. This is not an issue for our study which focuses on the instability stage prior to turbulent overturns.

In the 2D simulations, the bulk mixing coefficient is even lower after the transition to turbulence. This is not surprising because 2D dynamics does not allow for vortex stretching, which is essential for driving a forward cascade to dissipation scales. We have therefore decided to report $\Gamma_{bk}$ only for the 3D simulations.

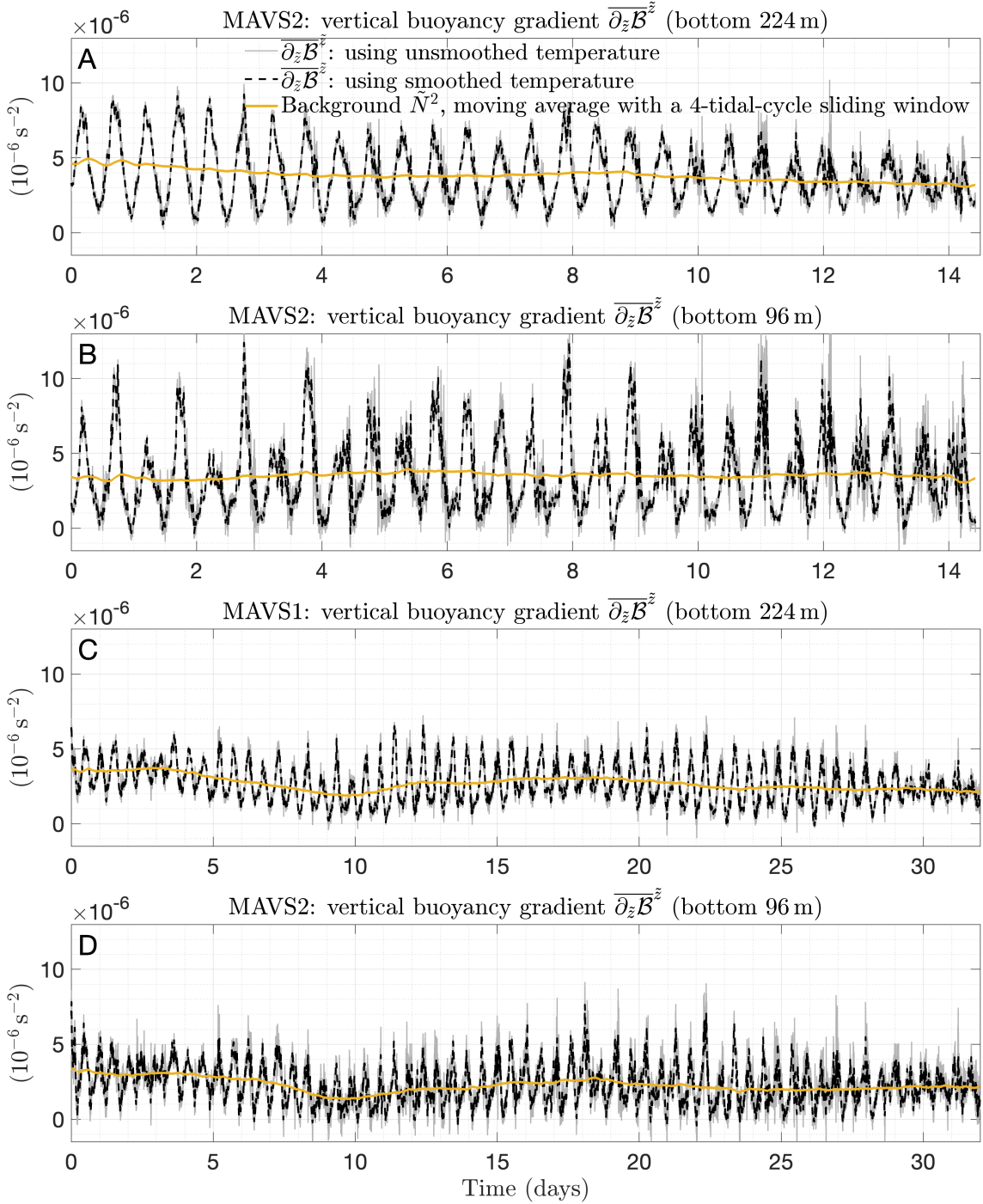

**Fig. S1.** Observed vertically averaged vertical buoyancy gradient from MAVS2 (*A*, *B*) and MAVS1 (*C*, *D*) moorings, covering the bottom 224 m (entire measurement depth; *A*, *C*) and the bottom 96 m (*B*, *D*). Gray lines show $\overline{\partial_{\tilde{z}}\mathcal{B}}^{\tilde{z}}$ computed using unsmoothed temperature data, with the linear relation between temperature and salinity (see Fig. 2*B*). To reduce noise in the observed vertical buoyancy gradient $\partial_{\tilde{z}}\mathcal{B}$, a Gaussian filter with a 15-min window size was applied to the temperature field. Black dashed lines indicate $\overline{\partial_{\tilde{z}}\mathcal{B}}^{\tilde{z}}$ calculated from the smoothed temperature data. The yellow lines represent the background $N^2$, calculated as the depth- and time-averaged vertical buoyancy gradient. The time average is computed using a moving average with a 4-M2-tidal-cycle sliding window (49 hours and 40 minutes). The $x-$axes of panels (*A*, *B*) represent days from 2021-07-07 to 2021-07-22. The $x-$axes of panels (*C*, *D*) represent days from 2021-08-01 to 2021-09-02.

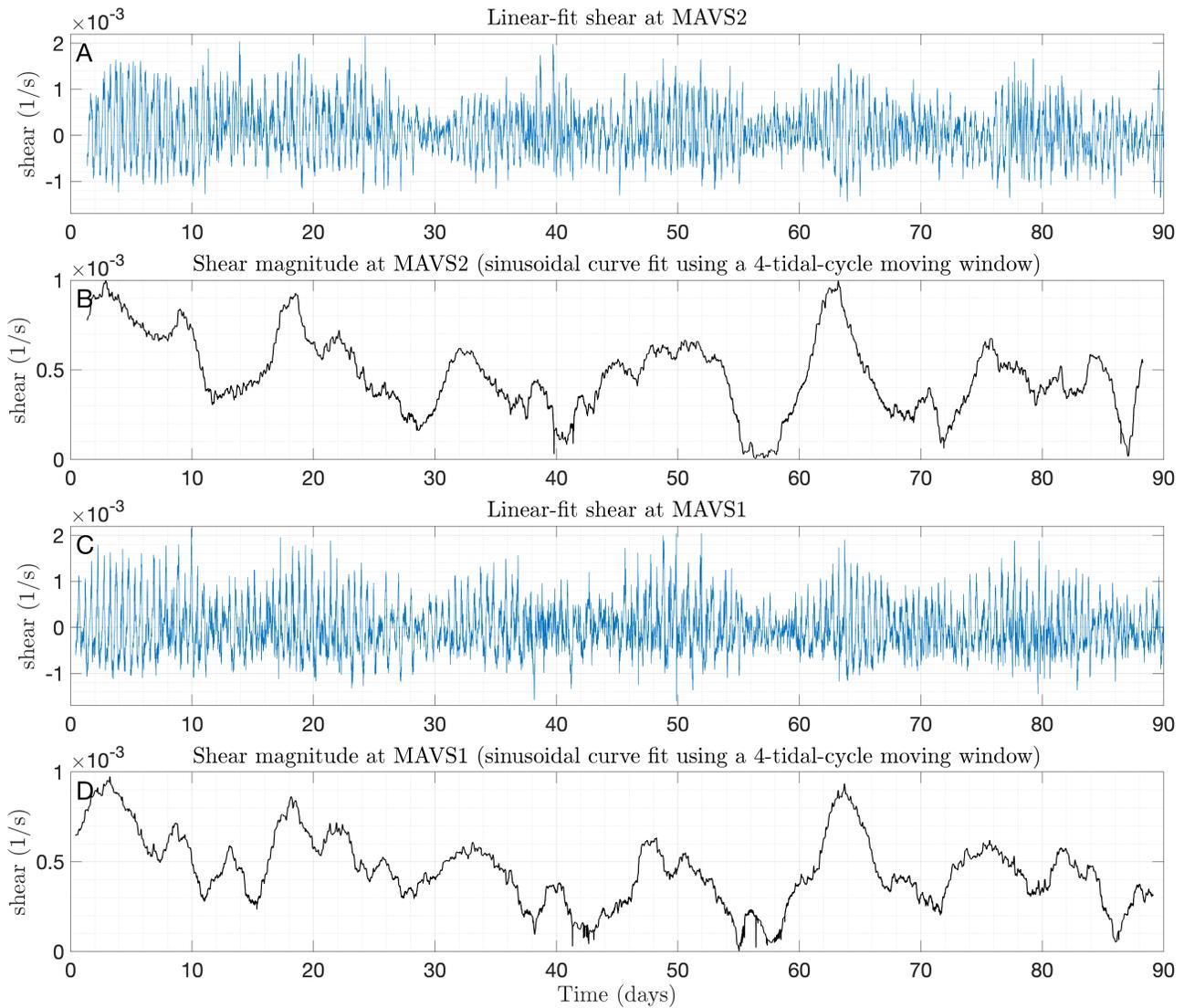

**Fig. S2.** Linear-fit slope-normal shear of the along-canyon velocity from the MAVS2 (*A*) and MAVS1 (*C*) moorings, spanning the full depth of measurement (bottom 224 m). Panels (*B*, *D*) show the shear magnitude at both moorings, calculated by applying sinusoidal curve fitting to the data in panels (*A*) and (*C*) using a moving window of four M2 tidal cycles (49 hours and 40 minutes). The $x-$axes represent days since 6 a.m. on 2021-07-06.

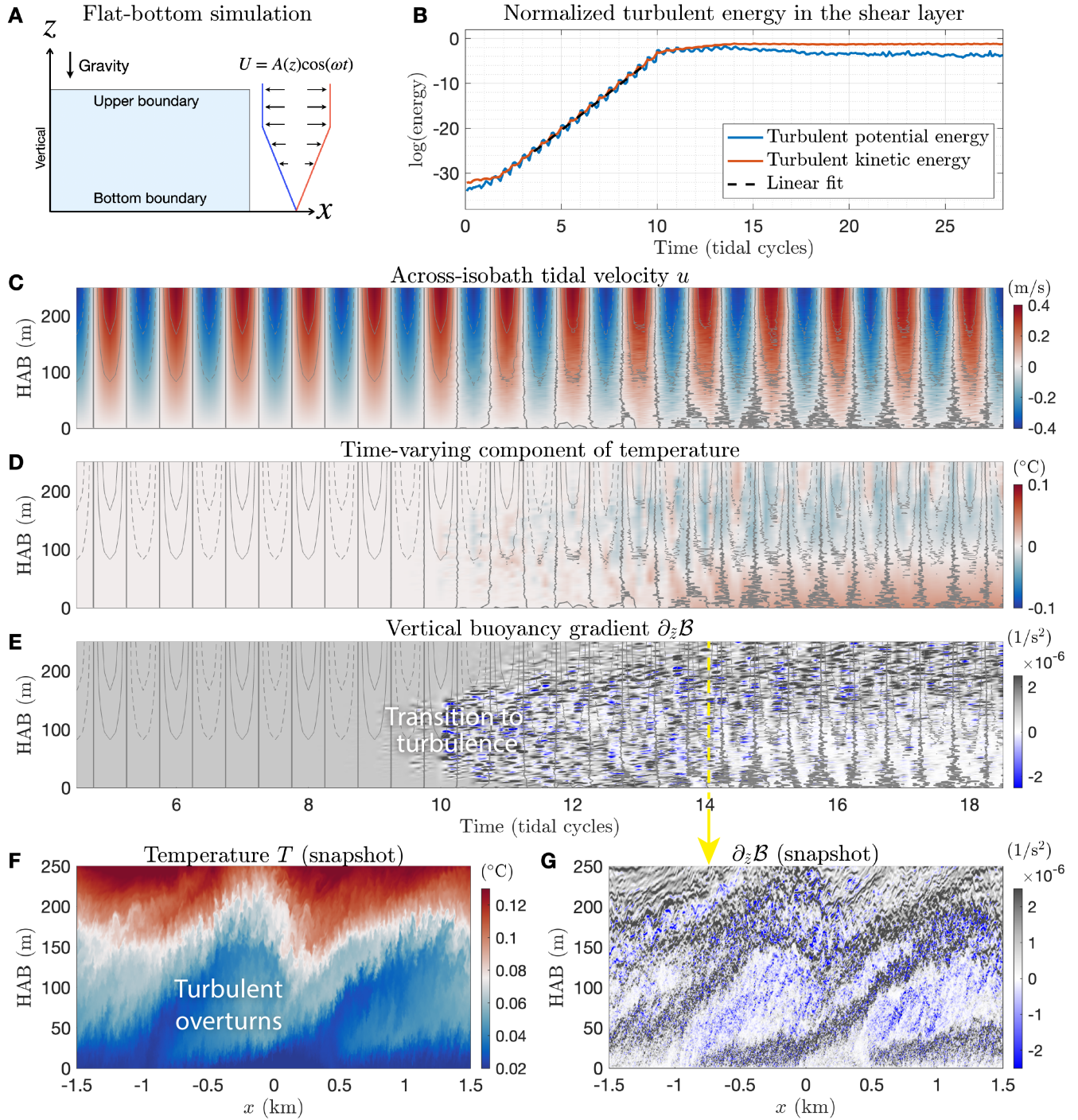

**Fig. S3.** Same as Fig. 6 in the main manuscript, but for the flat-bottom simulation (topographic slope $\theta = 0°$). The minimum Richardson number of the background flow is $Ri_{\min} = 0.31$. Note that the results remain qualitatively similar for Richardson numbers of order unity; a small value of $Ri_{\min}$ is chosen here to better visualize the turbulent overturns.

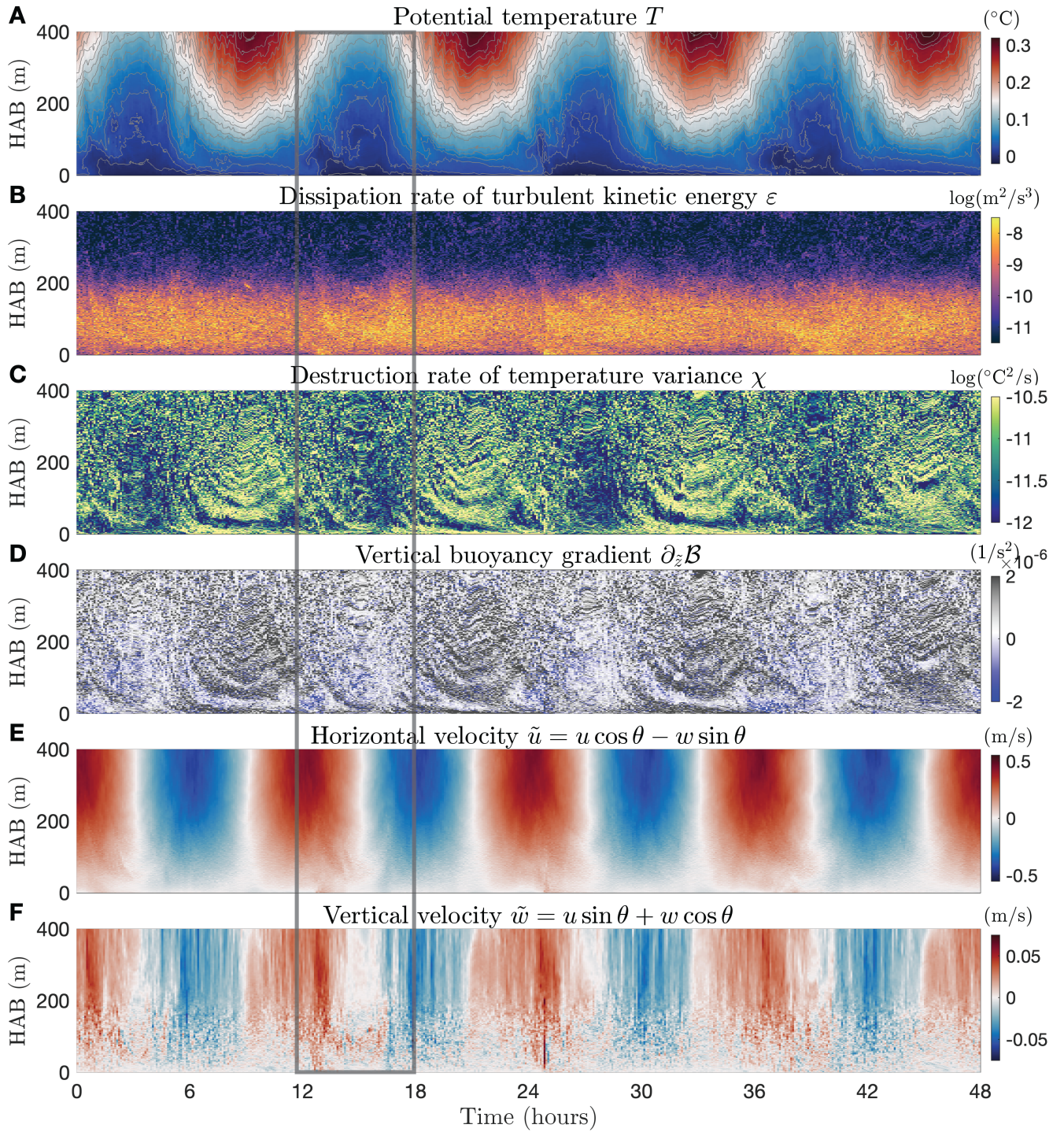

**Fig. S4.** Time series of the MITgcm simulation with a sloping topography ($\theta = 4°$) for the reference simulation (Fig. 6). The $x$−axes show hours since the beginning of the 17th tidal cycles and the $y$−axes represent height above thes bottom (HAB). Time series of (*A*) potential temperature, with thin gray contours indicating intervals of $0.015°$C, (*B*) dissipation rate of turbulent kinetic energy $\varepsilon$, (*C*) destruction rate of temperature variance $\chi$, (*D*) vertical buoyancy gradient, (*E*) horizontal velocity, and (*F*) vertical velocity. Panels *B*–*C* are plotted on a logarithmic scale with a base of 10. The time series were taken at a selected location ($x = 1$ km). The gray rectangle indicates the phase when the water column is most unstable. Note that, due to the idealization of the MITgcm simulation and the differences in hydrography, the magnitudes of $\varepsilon$ and $\chi$ computed from the model should not be directly compared to those observed in the BLT field campaign (2–4).

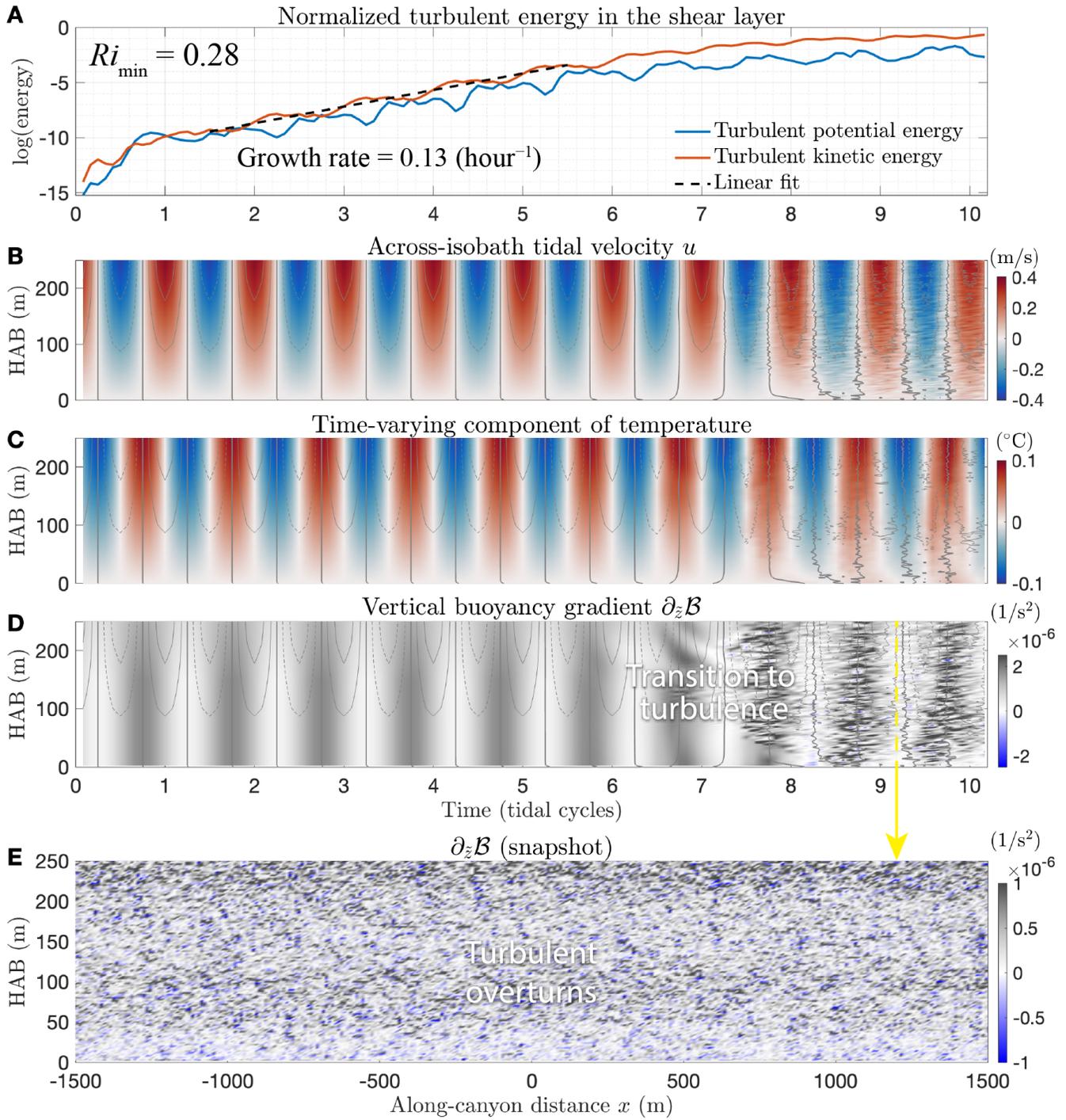

**Fig. S5.** Results of the 3D MITgcm simulation with rotation. The minimum Richardson number of the background tide is $Ri_{\min} = 0.28$. The topographic slope is $\theta = 4°$. The axis labels are the same as those in Fig. 6.

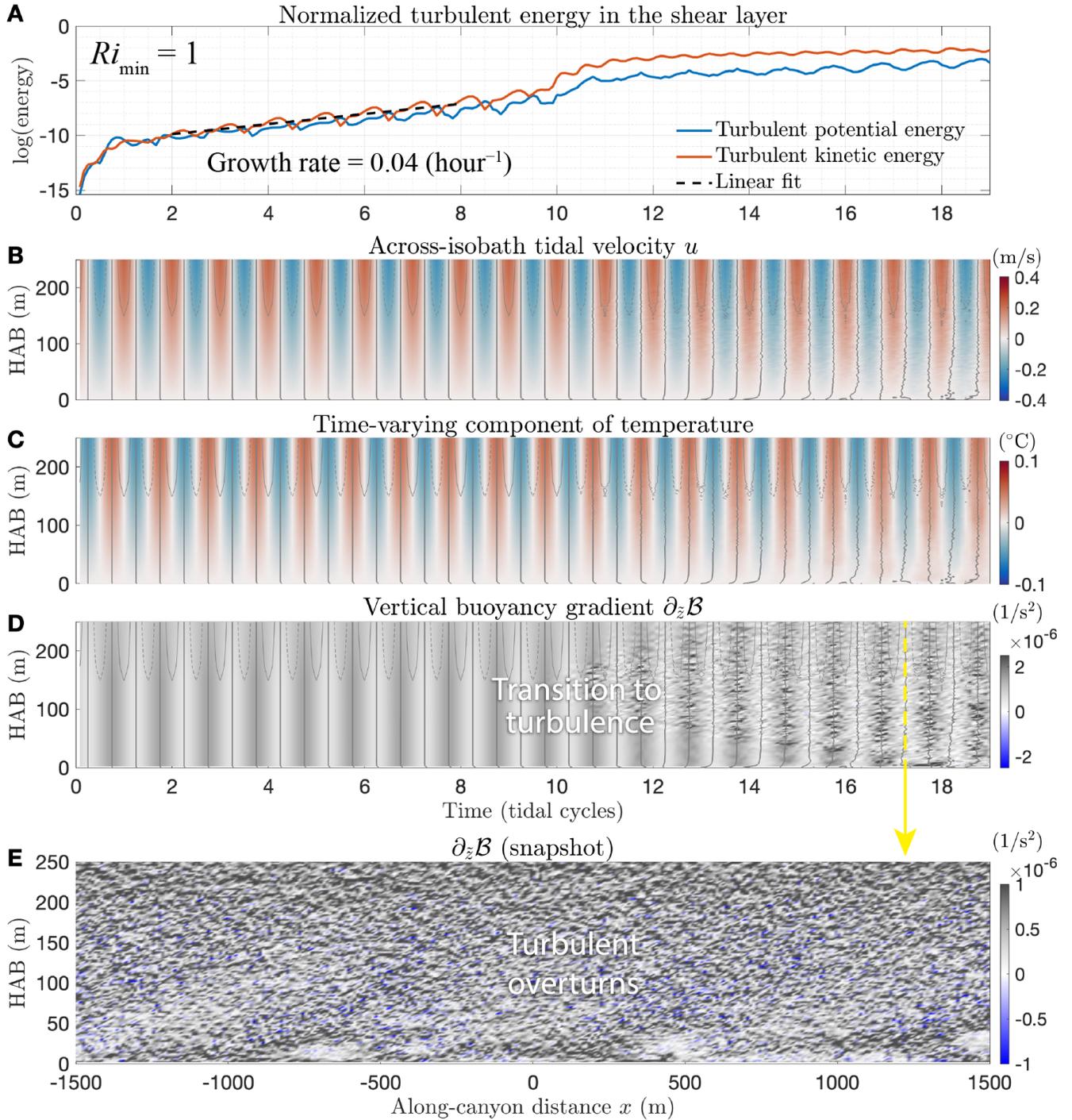

**Fig. S6.** Results of the 3D MITgcm simulation with rotation. The minimum Richardson number of the background tide is $Ri_{\min} = 1$. The topographic slope is $\theta = 4°$. The axis labels are the same as those in Fig. 6.

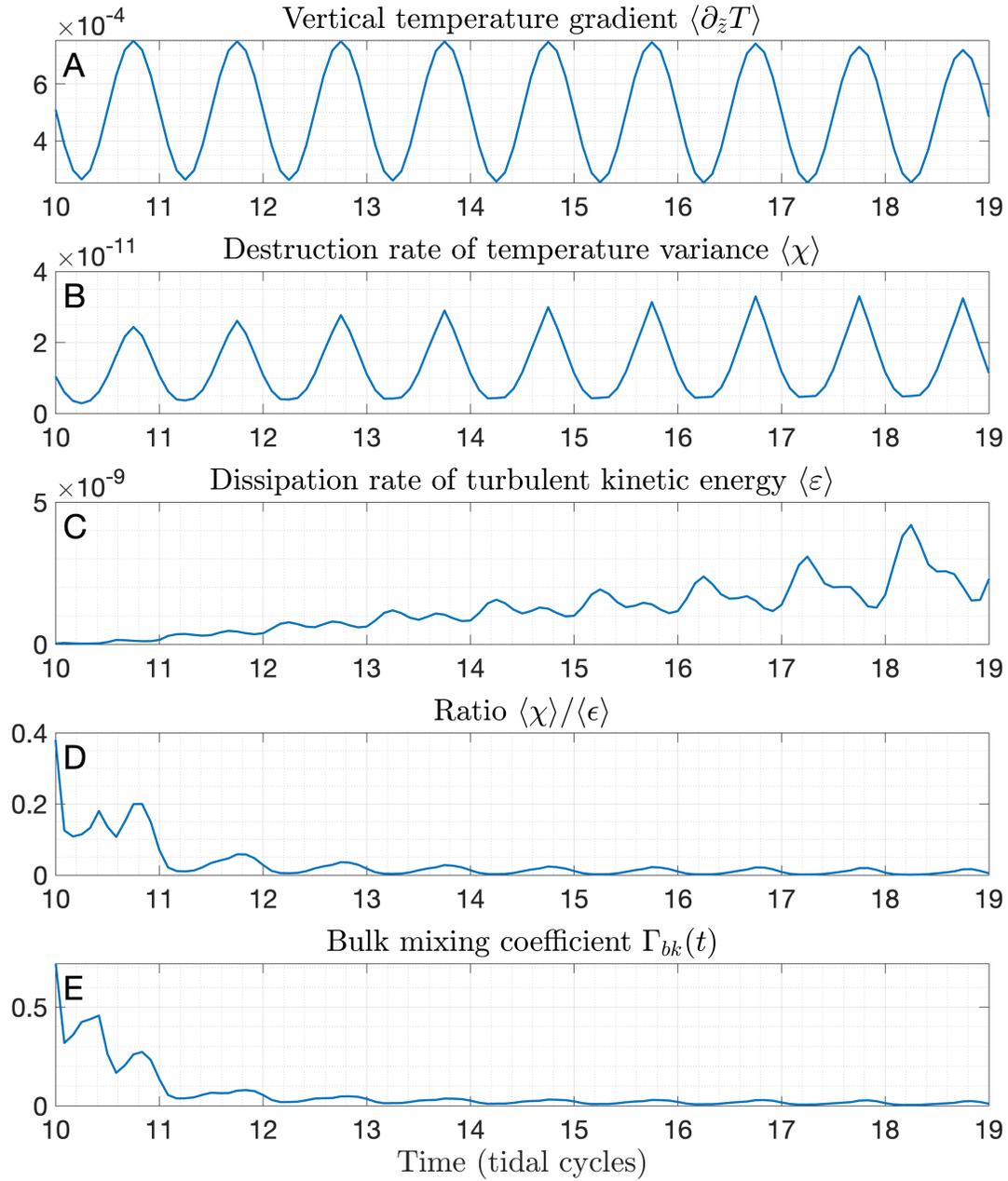

**Fig. S7.** Time series of volume-averaged (*A*) vertical temperature gradient, (*B*) the destruction rate of temperature variance, (*C*) the dissipation rate of turbulent kinetic energy, (*D*) the ratio $\langle\chi\rangle/\langle\varepsilon\rangle$, and (*E*) the bulk mixing coefficient. All computed from the 3D simulation with a $4°$ sloping bottom and $Ri_{\min} = 1$ (Fig. S6; Sec. 4 of the Supporting Information Text).

|    | Parameter | Value | Description |
|----|-----------|-------|-------------|
|    | $\theta$ | $0°, 4°$ | Topographic slope |
|    | $T_{\text{tide}}$ | $43200\,\text{s}$ | Tidal period |
|    | $\omega = 2\pi/T_{\text{tide}}$ | $1.4544 \times 10^{-4}\,\text{s}^{-1}$ | Tidal frequency |
|    | $N$ | $10^{-3}\,\text{s}^{-1}$ | Vertical background stratification |
|    | $\Lambda$ | $0 \sim 2 \times 10^{-3}\,\text{s}^{-1}$ | Background tidal shear amplitude |
|    | $L_x$ | $3,000\,\text{m}$ | Horizontal domain size |
|    | $H$ | $500\,\text{m}$ | Domain height for the simulations with quasi-linear tidal current amplitude (Fig. 5*B*) |
|    | $H$ | $800\,\text{m}$ | Domain height for the simulations with hyperbolic tangent tidal current amplitude (Fig. 5*C*) |
| 2D | $\Delta_x$ | $3\,\text{m}$ | Horizontal grid spacing |
|    | $\Delta_z$ | $1\,\text{m}$ | Vertical grid spacing |
|    | $\Delta_t$ | $0.5 \sim 10\,\text{s}$ | Time step (depending on the tidal amplitude) |
|    | $T_{\text{noise}}$ | $10^{-20}\,°\text{C}$ | Magnitude of the white noise used for initial temperature perturbation |
|    | $\alpha$ | $2 \times 10^{-4}\,°\text{C}^{-1}$ | Thermal expansion coefficient |
|    | $\nu$ | $5 \times 10^{-6}$ or $1 \times 10^{-5}\,\text{m}^2\text{s}^{-1}$ | Eddy viscosity for flat-bottom or sloping-bottom simulations |
|    | $\kappa$ | $5 \times 10^{-6}$ or $1 \times 10^{-5}\,\text{m}^2\text{s}^{-1}$ | Eddy diffusivity for flat-bottom or sloping-bottom simulations |
|    | $C_{\text{d}}$ | $2.5 \times 10^{-3}$ | Quadratic bottom-drag coefficient |
|    | $\theta$ | $4°$ | Topographic slope |
|    | $\Lambda$ | $1.0 \times 10^{-3}\,\text{s}^{-1}$ and $1.7 \times 10^{-3}\,\text{s}^{-1}$ | Background tidal shear amplitude |
|    | $L_x$ | $7,700\,\text{m}$ | Along-canyon domain size |
|    | $L_y$ | $2,400\,\text{m}$ | Cross-canyon domain size |
|    | $H$ | $500\,\text{m}$ | Domain height |
|    | $\Delta_x, \Delta_y$ | $10\,\text{m}$ | Horizontal grid spacing |
| 3D | $\Delta_z$ | $2\,\text{m}$ | Vertical grid spacing |
|    | $\Delta_t$ | $4.4 \sim 5.7\,\text{s}$ | Time step (depending on the tidal amplitude) |
|    | $f$ | $1.2 \times 10^{-4}\,\text{s}^{-1}$ | Coriolis parameter |
|    | $\nu$ | $2 \times 10^{-5}\,(4 \times 10^{-5})\,\text{m}^2\text{s}^{-1}$ | Horizontal (vertical) viscosity |
|    | $\kappa$ | $2 \times 10^{-5}\,(4 \times 10^{-5})\,\text{m}^2\text{s}^{-1}$ | Horizontal (vertical) diffusivity |

**Table S1. List of parameters used in the MITgcm simulations. For the 3D simulations, only those differing from the 2D case are listed.**



\end{document}